\documentclass[aps,prb,reprint,twocolumn,superscriptaddress,amsmath,amssymb,longbibliography,10pt]{revtex4-2}
\usepackage{graphicx}
\usepackage{epstopdf}
\usepackage{xcolor}
\usepackage{comment}
\usepackage[colorlinks, linkcolor=red,citecolor=blue,urlcolor=blue]{hyperref}
\usepackage[all]{hypcap} % let hyperlinks correctly point to figures rather than their captions;
\begin{document}
\title{ Coulomb screening of the orbital inverse Faraday effect in finite-size diffusive conductors}

\begin{abstract}
We study the effects of Coulomb screening on the second-order dc response in finite-size diffusive conductors. Starting from the Boltzmann equation supplemented by the self-consistent electrostatic field, we develop a kinetic theory of the inverse Faraday effect for a two-dimensional disk driven by a circularly polarized ac electric field. In this formulation, the first-order charge modulation and the induced electric field are coupled through the Coulomb interaction, and the self-consistent first-order solution determines the nonlinear source for the second-order dc circulating current responsible for the induced magnetic moment. The radial problem is solved numerically using an angular-harmonic expansion and the Chebyshev collocation method. It is shown that in the non-self-consistent approximation the frequency dependence of the second-order dc response has a maximum when the diffusion length associated with the radiation frequency becomes comparable to the sample radius. This behavior is found to be strongly modified by self-consistent screening. We demonstrate that Coulomb screening shifts the maximum of the dc response to a higher drive frequency and suppresses the peak amplitude. This behavior is observed consistently for the area-integrated circulating current and the induced magnetic moment. The obtained results clarify the limitations of the non-self-consistent description of the inverse Faraday effect and show that Coulomb self-consistency is crucial for quantitative predictions of optically induced dc magnetization in finite-size diffusive conductors.
\end{abstract}

\author{A. A. Kopasov}
\affiliation{National University of Science and Technology ``MISIS'', Moscow 119049, Russia}
\author{A. S. Mel'nikov}
\affiliation{Moscow Institute of Physics and Technology (National Research University), Dolgoprudnyi, Moscow region, 141701 Russia}
\affiliation{Institute for Physics of Microstructures, Russian Academy of Sciences, 603950 Nizhny Novgorod, GSP-105, Russia}

\maketitle

\section{Introduction}

Light-induced magnetization represents an important class of magneto-optical phenomena, in which an electromagnetic field induces a dc magnetic response in solids through nonlinear coupling to electronic, spin, or lattice degrees of freedom~\cite{Kimel2007_JPCM,Kirilyuk2010,Carva2017_HMM,Walowski2016_JAP}. One of the prominent effects of this type is the inverse Faraday effect (IFE), namely, the generation of a dc magnetic response by circularly polarized electromagnetic radiation. Since its original theoretical prediction by L.~P.~Pitaevskii~\cite{Pitaevskii1961} and the early experimental observation~\cite{vanDerZiel1965}, the IFE has been formulated phenomenologically as a nonlinear magneto-optical response whose sign is reversed when the polarization handedness is changed~\cite{Pershan1966,Pershan1967}. In magnetic media, the IFE provides a route by which circularly polarized femtosecond pulses can excite spin dynamics and contribute to ultrafast optical control of magnetic order~\cite{Kimel2005, Kimel2007_JPCM,Kirilyuk2010}.

The microscopic origin of the induced magnetization depends strongly on the electronic and collective degrees of freedom of the system. In insulating materials, the magnetic response can arise from virtual electronic transitions, optically driven phonons, dynamical multiferroicity, chiral lattice motion, or correlation effects in Mott systems~\cite{Nova2017,Juraschek2017,Juraschek2022,Luo2023,Banerjee2022,Gribova2024}. In conducting systems, circularly polarized radiation can generate dc circulating electronic currents. In systems with structural asymmetry and/or spin-orbit coupling, homogeneous illumination can generate a spin-dependent contribution to the IFE response~\cite{Edelstein1998,Tanaka2020,Xu2021,Tanaka2024}. The magnitude of this contribution is then controlled by the strength of spin-orbit coupling and by the details of the band structure. Irrespective of the presence of spin-orbit coupling, there is also an orbital contribution to the IFE, which is produced by rectified electric currents and is tied to spatial inhomogeneities. In finite samples such inhomogeneities are naturally supplied by boundaries, while in other settings they can originate from the edge of the illuminated region, impurities, or gradients of the electromagnetic field. The microscopic mechanisms and manifestations of the orbital IFE have been studied for a variety of normal conducting systems~\cite{Hertel2006,Hertel2015,Battiato2014,Tokman1999,Magarill1999,Quinteiro2009,Koshelev2015,Kibis2011,Kibis2013,Kibis2021,Tokman2020,Sharma2024,Kopasov2025Impurity} and in superconducting systems~\cite{Mironov2021_PRL_IFE_SCCondensates,Plastovets2022_PRB_AbrikosovVorticesIFE,Putilov2023_JETPL_IFE_FiniteGap,Plastovets2023_PhysLettA_FluctuationIFE_Rings,Dzero2024_PRB_IFE_DisorderSC}.

Many standard descriptions of the orbital IFE in conducting systems are focused on the high-frequency regime and mainly address the response of the electronic system to a prescribed electromagnetic field~\cite{Hertel2006,Hertel2015,Battiato2014,Tokman2020,Sharma2024}. Useful reference point for the discussion of the orbital IFE in conducting systems is provided by a Drude-type generalization of the Hertel theory~\cite{Hertel2006,Hertel2015}, in which disorder is included through the elastic momentum relaxation time $\tau$. For circularly polarized radiation with frequency $\omega$ and electric-field complex amplitude $\mathbf{E}_{\omega}$, this theory gives the magnetic moment
\begin{equation}\label{magnetic_moment_Hertel}
\mathbf{M} \propto V\frac{\tau^2}{\omega(1 + \omega^2\tau^2)}i(\mathbf{E}_{\omega}\times\mathbf{E}_{\omega}^*) \ .
\end{equation}
Here $V$ is the volume of the sample. Although several routes to enhanced IFE response have been discussed, including plasmonic effects and synthetic magnetic fields~\cite{Potashin2020,Han2023,Durnev2023}, the above expression naturally suggests the enhancement of the induced dc magnetic moment as the radiation frequency is decreased. However, the actual low-frequency behavior of the IFE response cannot be inferred from the above estimate alone. Indeed, in the dirty limit $\omega\tau\ll1$, Eq.~(\ref{magnetic_moment_Hertel}) becomes singular, $\mathbf M\propto1/\omega$. Despite significant progress in understanding orbital IFE mechanisms in metals, mesoscopic structures, plasmonic systems, and superconductors, the low-frequency regime in finite normal conductors remains less understood, especially when finite-size charge relaxation and Coulomb screening have to be treated self-consistently.

There are two reasons why the low-frequency regime of orbital IFE requires a detailed analysis. First, Eq.~(\ref{magnetic_moment_Hertel}) gives only the integral magnetic moment and does not resolve the spatial distributions of the induced charge and current in the sample. These distributions involve intrinsic length scales such as $v_F/\omega$ in the ballistic limit~\cite{Bespalov2025}, the diffusion length $\sqrt{D/\omega}$ in the diffusive regime, and the Thomas--Fermi screening length $q_{\rm TF}^{-1}$. Here $v_F$ is the Fermi velocity and $D$ is the diffusion coefficient, and $q_{\rm TF}$ is the Thomas--Fermi screening wave number. At sufficiently low frequencies, the diffusive charge redistribution generated by the ac field can extend over distances comparable to the sample size, so that the magnetic response can be determined by sample-wide profiles of the induced charge and current. Existing studies of related phenomena, including surface photogalvanic effects, photon drag, edge photocurrents, and structured-light-induced currents, demonstrate that boundaries, disorder, and field inhomogeneity can qualitatively modify second-order transport responses~\cite{Ivchenko2005,MagarillEntin1981,Alperovich1981,Gurevich1993,Mikheev2018,Karch2011,Plank2019,Candussio2020,DurnevTarasenko2019,DurnevTarasenko2020,Gunyaga2023,Gunyaga2025,Bespalov2025,Tanaka2023_NoncentrosymmetricSC,Vaswani2020_THzSHG_SC,Nakamura2020_NbN_SHG,Itahashi2022_SecondHarmonicSC,Tanaka2024_NonlinearOpticalSC}. 
The second reason is related to Coulomb screening. In the non-self-consistent treatment, the response is calculated for a prescribed electromagnetic field, while the electric field generated by the induced charge redistribution is neglected. Such feedback of the internal potential is known to be essential in dynamic and nonlinear transport, where it is required for charge conservation, current conservation, and gauge invariance~\cite{Buttiker1993,ChristenButtiker1996,Christen1997}. In the present finite diffusive problem this issue becomes especially relevant at low frequencies, where the charge redistribution can extend over distances comparable to the sample size. Since screening modifies both the local electric field acting on the electrons and the induced charge distribution entering the nonlinear source, its net effect on the magnetic response cannot be inferred from the non-self-consistent theory alone. To the best of our knowledge, a self-consistent theory of the orbital IFE in finite-size conductors has not been developed. The main goal of this work is to fill this gap and determine how Coulomb screening modifies the orbital IFE in a finite diffusive conductor.

\begin{figure}[!htpb]
\centering
\includegraphics[scale=0.5]{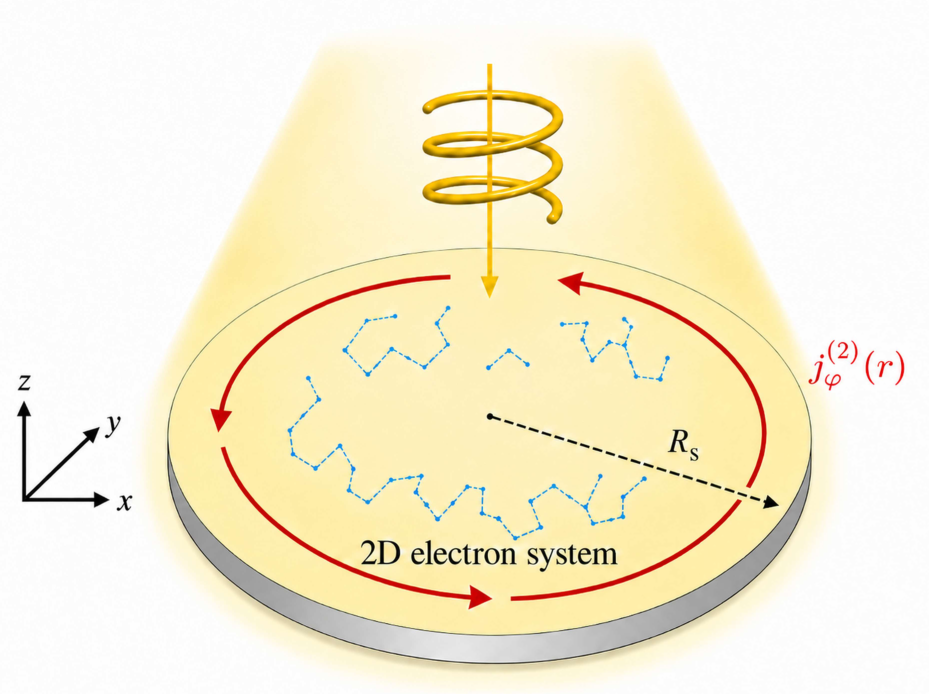}
\caption{Schematic picture of the model system. A finite two-dimensional electron system in the form of a disk of radius $R_s$ is irradiated by normally incident circularly polarized radiation. The red arrows show the induced second-order dc circulating current density $j_{\varphi}^{(2)}(r)$. Blue dashed trajectories illustrate diffusive electron motion due to impurity scattering.}
\label{model_system}
\end{figure}

In this paper we study the effects of Coulomb screening on the second-order dc response in finite diffusive conductors. Starting from the Boltzmann equation supplemented by the self-consistent electrostatic field, we develop a kinetic theory of the inverse Faraday effect for a two-dimensional disk of radius $R_s$ driven by a circularly polarized ac electric field (see Fig.~\ref{model_system}). In this formulation Coulomb interaction enters already at the level of the linear ac response: the first-order charge modulation generates an induced electric field, which acts back on the electronic distribution. The resulting self-consistent first-order solution then determines the nonlinear source for the angular-independent second-order dc circulating current responsible for the induced magnetization. The radial problem is solved numerically using an angular-harmonic expansion and the Chebyshev collocation method~\cite{Trefethen2000_SpectralMethodsMATLAB}.

We first analyze the non-self-consistent approximation, where the frequency dependence of the second-order dc response has a maximum when the diffusion length associated with the ac drive becomes comparable to the sample radius, i.e., at the Thouless frequency scale $\omega \sim D/R_s^2$. We then show that such behavior is strongly modified by self-consistent screening of the first-order charge modulation. It is demonstrated that Coulomb screening shifts this maximum to a higher frequency by an amount of order $\delta \omega \sim Dq_{\rm TF}/R_s$ and suppresses the peak amplitude. This behavior is observed consistently for the area-integrated circulating current and for the induced magnetic moment. The obtained results clarify the limitations of the non-self-consistent description of the inverse Faraday effect and show that Coulomb self-consistency is crucial for quantitative predictions of optically induced dc magnetization in finite diffusive conductors.

The pronounced dependence of the inverse Faraday response on screening suggests that this effect can be used to probe charge dynamics in mesoscopic conductors. In particular, measurements of the position and magnitude of the response maximum as functions of frequency and sample size can provide information about characteristic charge-relaxation and screening scales. More generally, the developed theoretical framework provides a natural route to study the role of Coulomb screening in other nonlinear response phenomena, including the surface photogalvanic effect and the photon drag.

The rest of the paper is organized as follows. Sec.~\ref{sec:basic_equations} introduces the model, the kinetic equation, and the self-consistent treatment of the induced electric field. Sec.~\ref{sec:non_self_consistent} discusses the non-self-consistent approximation and the maximum in the frequency dependence of the inverse Faraday response. Sec.~\ref{sec:self_consistent_response_section} develops the self-consistent kinetic formulation of the linear ac response and the second-order dc response. Sec.~\ref{sec:numerical_results} presents the numerical results and discusses the Coulomb-induced suppression and frequency shift of the inverse Faraday response. The results are summarized in Sec.~\ref{sec:summary}.

\section{Basic equations}
\label{sec:basic_equations}

We consider a degenerate two-dimensional electron gas with quadratic energy-momentum relation. The sample is patterned in the form of a disk of radius $R_s$, and the electron motion is confined to the $xy$ plane (see Fig.~\ref{model_system}). 
The system is irradiated by normally incident circularly polarized electromagnetic radiation. We assume that the local ac electric field driving the electron motion is spatially homogeneous in the plane of the disk. Possible effects of the magnetic field of the incident wave on the electron motion are neglected.
Corresponding corrections associated with the ac magnetic field should be small in the limit $v_F/c \ll 1$, where $v_F$ is the Fermi velocity and $c$ is the speed of light. We use Gaussian units and set $\hbar=1$. The external electric field reads as
\begin{equation}
    \mathbf E^{\rm ext}(t) = \mathbf E^{\rm ext}_\omega e^{-i\omega t}+\mathbf E_\omega^{{\rm ext}*} e^{i\omega t} \ .
    \label{eq:E_real}
\end{equation}
Here $\mathbf E^{\rm ext}_{\omega}$ denotes the local ac field that drives the electronic motion in the absence of the longitudinal electric field induced by charge redistribution. It can be identified with the incident-wave amplitude when reflection from the conducting layer and its environment is negligible. More generally, $\mathbf E^{\rm ext}_{\omega}$ should be understood as the actual tangential field at the conducting layer. In polar coordinates $(r, \varphi)$ with the origin at the center of the disk, the projections of the complex amplitude of the field can be written in the form 
\begin{subequations}
\label{helicity_decompositions}
\begin{align}
    E_{\omega, r}^{\rm ext} = E_- e^{i\varphi} + E_+ e^{-i\varphi} \ , \ \ \ \ \ \\
    E_{\omega, \varphi}^{\rm ext} = i\left(E_- e^{i\varphi}-E_+ e^{-i\varphi}\right) \ ,
\end{align}
\end{subequations}
and the circular components of the field are defined as $E_\pm = (E_x\pm iE_y)/2$.

The dynamics of the electronic system is treated within the kinetic equation approach. In the present work we focus on the effects of elastic impurity scattering described by a relaxation-time collision integral which relaxes the anisotropic part of the distribution function. For simplicity we disregard inelastic relaxation. This approximation is appropriate when the response  is mainly controlled by diffusion during one period of the drive. More specifically, this work addresses the following hierarchy of length scales and parameters:
\begin{equation}
p_F^{-1} \ll \ell \ll R_s \ll L_{\rm in} \ , \ \ \ \omega\tau \ll 1 \ ,
\end{equation}
where $p_F$ is the Fermi momentum, $\ell = v_F\tau$ is the elastic mean-free path, $\tau$ is the elastic momentum relaxation time, $L_{\rm in} = \sqrt{D\tau_{\rm in}}$ is the distance over which electrons diffuse during the inelastic relaxation time $\tau_{\rm in}$, and $D = v_F^2\tau/2$ is the two-dimensional diffusion coefficient.

Under the employed assumptions for electron dynamics, the Boltzmann equation for the distribution function $f$ takes the form 
\begin{equation}
    \left(\partial_t + \mathbf v\cdot\nabla_{\mathbf r} + e\mathbf E(\mathbf r,t)\cdot\nabla_{\mathbf p}\right)
    f(\mathbf r,\mathbf p,t) = -\frac{f-\langle f\rangle_{\mathbf{p}}}{\tau} \ . 
    \label{eq:boltzmann}
\end{equation}
Here $\mathbf v=v\hat{\mathbf p}$ is the particle velocity~\footnote{The momentum dependence is kept in Eq.~(\ref{eq:boltzmann}) because the second-order source terms contain derivatives with respect to $\mathbf{p}$.
After the energy integration in the degenerate limit, the reduced kinetic
equations are evaluated at the Fermi surface.}, $\hat{\mathbf{p}} = \mathbf{p}/p$ is the unit vector in the direction of the  momentum, $e = -|e|$ is the electron charge, and $\langle\cdots\rangle_{\mathbf{p}}$ denotes angular averaging over the momentum direction.
The collision integral in Eq.~(\ref{eq:boltzmann}) describes elastic scattering by point impurities. Such scattering randomizes the direction of the electron momentum while conserving the single-particle energy and therefore relaxes only the angular dependence of the distribution function on a constant-energy surface~\cite{RammerSmith1986}. Relaxation toward a local equilibrium distribution requires inelastic processes, which are neglected in the present work.
 The induced charge density and the current density are defined as
\begin{subequations}
\label{charge_current_definitions}
\begin{align}
   \label{charge_density_definition}
   \delta \rho(\mathbf r,t) = 2e \int \frac{pdp}{2\pi} \left\langle \delta f(\mathbf r,\mathbf p,t)\right\rangle_{\mathbf{p}} \ ,\\
   \label{current_density_definition}
    \mathbf{j}(\mathbf{r},t) = 2e \int \frac{pdp}{2\pi} \left\langle \mathbf{v} f(\mathbf r,\mathbf p,t) \right\rangle_{\mathbf{p}} \ .
 \end{align}
\end{subequations}
In the above expressions $\xi = p^2/2m^* - \mu$, $m^*$ is the effective mass, $\mu$ is the chemical potential, $\nu = m^*/2\pi$ is the density of states per one spin projection at the Fermi level, $\delta f(\mathbf{r},\mathbf{p},t) = f(\mathbf{r},\mathbf{p},t) - f_0(\xi)$, $f_0(\xi)$ is the Fermi-Dirac distribution function. The temperature is set to zero, so that $\partial f_0/\partial \xi = -\delta(\xi)$ and $\delta(x)$ is the Dirac delta function.

The self-consistency considered in this work is electrostatic in nature.  
Under this assumption, the charge redistribution inside the sample produces an additional longitudinal electric field.
Thus, the electric field entering the Boltzmann equation~\eqref{eq:boltzmann} is the sum of the external field and the self-consistent electrostatic field generated by the
induced charge,
\begin{equation}
    \mathbf E(\mathbf r,t) = \mathbf E^{\rm ext}(t) + \mathbf E^{\rm ind}(\mathbf r,t) \ , \ \ \ 
    \mathbf E^{\rm ind}=-\nabla\Phi \ .
    \label{eq:total_field}
\end{equation}
It is important to note that this formulation does not include the transverse electromagnetic self-field generated by the induced ac current. The corresponding nonretarded approximation is controlled by $\omega R_s/c\ll1$ for sample-wide charge profiles and by $\sqrt{D\omega}/c\ll1$ when the response is localized on the diffusion length. The latter condition, which can be rewritten as $\omega\tau \ll 2 (c/v_F)^2$, is automatically satisfied in the diffusive limit $\omega\tau \ll 1$. Thus, the restrictive condition posed by both the diffusive limit and nonretarded approximation is as follows: $\omega \ll {\rm min}(\tau^{-1}, \ c/R_s)$.

The potential $\Phi(\mathbf{r},t)$ in Eq.~(\ref{eq:total_field}) is obtained from the solution of the three-dimensional Poisson problem $\Phi_{\rm 3D}(\mathbf{r},z,t)$ which is then projected onto the conducting layer. It is important to note that $\Phi_{\rm 3D}$ depends on the environment surrounding the conducting layer. Here we consider an isolated disk embedded in a homogeneous static dielectric medium and, for brevity, set the dielectric constant $\epsilon_{\rm env}$ to unity in the expressions below. A constant dielectric background can be restored by replacing the Coulomb functional according to $\Phi[\delta \rho] \to \Phi[\delta\rho]/\epsilon_{\rm env}$. More complicated environments, including substrates, dielectric multilayers, or nearby gates, modify the Coulomb kernel and can produce a momentum- and frequency-dependent effective interaction. These effects lie outside the scope of the present model. Introducing the normalized transverse distribution of the electron density $w(z)$ ($\int dz\, w(z)=1$), the three-dimensional electrostatic potential satisfies the Poisson equation
\begin{equation}
    \left(\nabla_{\mathbf{r}}^2 + \partial_z^2 \right) \Phi_{\rm 3D}(\mathbf r,z,t) = -4\pi \,\delta \rho(\mathbf r,t)w(z) \ ,
    \label{eq:poisson_3d}
\end{equation}
and the effective potential acting on the two-dimensional electrons is its transverse average
\begin{equation}
    \Phi(\mathbf r,t) = \int dz \ w(z)\Phi_{\rm 3D}(\mathbf r,z,t) \ .
    \label{eq:phi_projected}
\end{equation}
For a disk geometry it is convenient to perform the angular decomposition
\begin{subequations}
\begin{align}
    \label{eq:angular_expansion_rho}	
    \delta \rho(\mathbf r,t) = \sum_m \delta \rho_m(r, t)e^{im\varphi} \ ,\\
    \label{eq:angular_expansion_Phi}	
     \Phi(\mathbf r,t) = \sum_m \Phi_m(r, t)e^{im\varphi} \ ,\\
     \label{eq:angular_expansion_Phi_3d}	
     \Phi_{\rm 3D}(\mathbf{r},z,t) = \sum_m \Phi_{\rm{3D}, m}(r,z,t)e^{im\varphi} \ ,
\end{align}
\end{subequations}
and solve Eq.~\eqref{eq:poisson_3d} by a Hankel transform, which yields (see Appendix~\ref{app:coulomb_closure} for the derivation)
\begin{eqnarray}
    \label{eq:phi_m_hankel}
    \Phi_m(r, t) = 2\pi  \int_0^\infty d\kappa \int_0^{R_s} dr' r' F(\kappa) \\
    \nonumber
    \times J_m(\kappa r)J_m(\kappa r')\delta \rho_m(r', t) \ .
\end{eqnarray}
Here $J_m(x)$ denotes the Bessel function of the first kind of order $m$~\cite{AbramowitzStegun1964}, and the finite-thickness form factor $F(\varkappa)$ has the following form:
\begin{equation}
    F(\kappa) = \int dz dz' w(z)w(z') e^{-\kappa |z-z'|}.
    \label{eq:form_factor_general}
\end{equation}
For definiteness, we choose the uniform-slab form factor $w(z) = 1/d$ ($-d/2<z<d/2$) for the calculations, where $d$ is the thickness of the 2D system. The corresponding expression has the form
\begin{equation}
    F(\kappa) = \frac{2}{d^2\kappa^2}\left(-1+\kappa d+e^{-\kappa d}\right) \ .
    \label{eq:form_factor_slab}
\end{equation}
The strict two-dimensional limit is recovered by taking $w(z) = \delta(z)$, for which $F(\varkappa) = 1$. For the finite-thickness profile in Eq.~(\ref{eq:form_factor_slab}) one has $F(\varkappa) \approx 1 - \varkappa d/3$ for $\varkappa d \ll 1$. Near the maximum of the inverse Faraday response the dominant finite-size charge mode has a characteristic wave vector $q_* \sim 1/R_s$. Since $d\ll R_s$ in the considered case, one gets $q_* d \ll 1$, so that finite thickness doesn't qualitatively modify the screening-induced shift and suppression of the response maximum. At higher frequencies, larger wave-vector components become increasingly important and finite-thickness effects can be more pronounced.

Note that the factorized form of $\delta\rho_{\rm 3D}$ in Eq.~(\ref{eq:poisson_3d}) implies the assumption that the transverse charge profile of the conducting layer is fixed. In other words, the form factor~(\ref{eq:form_factor_general}) describes the finite-thickness geometry for the Coulomb interaction, but does not account for a dynamical redistribution of charge density across the thickness. This approximation is natural for quantum-confined two-dimensional electron systems, where $w(z)$ is fixed by the transverse part of the wave function. For thin conducting films it represents a quasi-two-dimensional description, valid when transverse charge redistribution can be neglected or absorbed into an effective static form factor. In thicker metallic films, where the induced charge may rearrange substantially in the transverse direction, the effective Coulomb kernel has to be obtained from the corresponding three-dimensional electrostatic screening problem.

Employing the angular decomposition of the induced electric field 
\begin{equation}
\mathbf{E}^{\rm ind}(\mathbf{r},t) = \sum_m \mathbf{E}_m^{\rm ind}(r,t)e^{im\varphi} \ ,
\end{equation}
the projections of the induced field for the angular harmonic $m$ are then
\begin{subequations}
    \label{eq:Eind_general_m}
\begin{align}
    E^{\rm ind}_{r,m}(r,t) = -\frac{\partial \Phi_m(r,t)}{\partial r} \ ,\\
    E^{\rm ind}_{\varphi,m}(r,t) = -\frac{im}{r}\Phi_m(r,t) \ .
    \end{align}
\end{subequations}

Equations~\eqref{eq:boltzmann} and (\ref{eq:total_field}) are solved perturbatively in the amplitude of the
external ac field,
\begin{equation}
    f=f_0+\delta f \ , \ \ \delta f=f^{(1)}+f^{(2)}+\cdots .
\end{equation}
The first-order correction for the distribution function $f^{(1)}$ oscillates at the radiation frequency $\omega$, while the second-order correction $f^{(2)}$ generally contains both a component at $2\omega$ and a time-independent component. The inverse Faraday response considered in this work is associated with the angular-independent part of the second-order dc current, which we write as
\begin{equation}
    \mathbf j_{\rm dc}^{(2)}(\mathbf r) = j_\varphi^{(2)}(r)\,\mathbf e_\varphi + \ldots \ .
    \label{inverse_Faraday_current}	
\end{equation}
The dots on the right-hand side of the above expression denote other angular components of the second-order dc response, which do not contribute to the net orbital magnetic moment. In particular, the full second-order dc response can also contain radial rectified currents, which are important in related photogalvanic problems with boundaries or spatially structured radiation~\cite{Gunyaga2023}. In the present work we focus on the orbital inverse Faraday response, which is determined only by the angular-independent azimuthal current $j_\varphi^{(2)}(r)$. The projection of the induced magnetic moment on the direction perpendicular to the plane of the disk is
\begin{equation}
    M_z = \frac{1}{2c} \int d^2\mathbf{r}\left[\mathbf r\times \mathbf j_{\rm dc}^{(2)}(\mathbf r)\right]_z =
    \frac{\pi}{c}\int_0^{R_s}dr \, r^2 j_\varphi^{(2)}(r) \ ,
    \label{eq:Mz_def}
\end{equation}
and we also calculate the area-integrated circulating current,
\begin{equation}
    I_{\rm area} = 2\pi \int_0^{R_s}dr\,r j_\varphi^{(2)}(r) \ ,
    \label{eq:Iarea_def}
\end{equation}
as an additional measure of the total dc circulating response. The magnetic moment $M_z$ is the primary observable associated with the inverse Faraday effect, whereas $I_{\rm area}$ characterizes the integrated azimuthal current density in the disk. Hereafter, $\mathbf j_{\rm dc}^{(2)}$ is understood as a two-dimensional sheet current density. Correspondingly, $M_z$ denotes the total orbital magnetic moment generated by this sheet current distribution.

\section{Non-self-consistent approximation}
\label{sec:non_self_consistent}

Let us discuss the non-self-consistent approximation. In this approximation the electric field is identified with the external one $\mathbf E(\mathbf r,t)\rightarrow \mathbf E^{\rm ext}(t)$. This approximation provides a useful reference problem because it already contains the finite-size diffusion mode responsible for the enhancement of the orbital inverse Faraday response. We first solve the linear problem, because the second-order dc current will be expressed through the first-order charge modulation.

Considering the linear response, we write the first-order correction for the distribution function in the form
\begin{equation}
\label{eq:nsc_f1_expansion}
f^{(1)}(\mathbf{r},\mathbf{p},t) = f_{\omega}^{(1)}(\mathbf{r},\mathbf{p})e^{-i\omega t} + {\rm c.c.} \ .
\end{equation}
In the diffusive regime the angular dependence of the distribution function can be truncated at the isotropic component and the first angular harmonic in momentum space
\begin{equation}
    f_\omega^{(1)}(\mathbf{r},\mathbf{p}) = f_{\omega,0}^{(1)}(\mathbf{r},\xi) + \hat{\mathbf p}\cdot \mathbf f_{\omega, 1}^{(1)}(\mathbf{r},\xi) \ .
    \label{eq:nsc_f1_truncation}
\end{equation}
To leading order in the diffusive expansion, the equations for these two harmonics reduce, after integration over the quasiparticle energy, to the continuity equation and the current density (see Appendix~\ref{app:diffusive_reduction} for details of the derivation)
\begin{subequations}
\label{continuity_current_nsc}
\begin{align}
    \label{eq:nsc_continuity}
    -i\omega \delta\rho_\omega^{(1)} + \nabla_{\mathbf r}\cdot \mathbf j_\omega^{(1)} = 0 \ ,\\
    \label{eq:nsc_diffusive_current}
    \mathbf j_\omega^{(1)} = -  D \nabla_{\mathbf r}\delta\rho_\omega^{(1)} + \sigma \mathbf E_\omega^{\rm ext} \ .
\end{align}
\end{subequations}
Here $\sigma = 2e^2\nu D$. Thus, the current density is the sum of the Drude response and the diffusion current. Due to the fact that the external electric field is spatially homogeneous in the disk $\nabla_{\mathbf r}\cdot\mathbf E_\omega^{\rm ext}=0$, the first-order charge modulation is generated by the boundary condition that the radial current vanishes at the sample edge,
\begin{equation}
    j_{\omega, r}^{(1)}(r = R_s,\varphi)=0 \ .
    \label{eq:nsc_bc_no_radial_current}
\end{equation}
The above condition expresses the absence of particle flux through the disk edge. In the full kinetic formulation of Sec.~\ref{sec:self_consistent_response_section}, the same condition follows from the specular reflection condition. This should be distinguished from the boundary condition issue discussed in Ref.~\cite{BrownDeng2021}. In that work the physical current is defined to vanish outside the metallic slab, while a finite normal component of the current at the interface doesn't have to be set to zero separately.  
Using the decomposition of the external field~(\ref{helicity_decompositions}), the first-order charge density can be written as
\begin{equation}
    \delta\rho_\omega^{(1)}(r,\varphi) = E_-\chi_\omega(r)e^{i\varphi} + E_+\chi_\omega(r)e^{-i\varphi} \ ,
    \label{eq:nsc_rho_decomposition}
\end{equation}
with both angular sectors ($\propto e^{\pm i\varphi}$) described by the same radial function $\chi_\omega(r)$. Substituting the above expansion into Eqs.~(\ref{continuity_current_nsc}), one obtains the Bessel equation
\begin{equation}
    -i\omega\chi_\omega - D \left(\frac{1}{r} \frac{d}{dr} r\frac{d}{dr} - \frac{1}{r^2} \right) \chi_\omega = 0 \ ,
    \label{eq:nsc_chi_equation}
\end{equation}
and Eq.~\eqref{eq:nsc_bc_no_radial_current} gives
\begin{equation}
    \frac{d\chi_\omega}{dr}
    \biggl|_{r=R_s} = 2e^2\nu \  .
    \label{eq:nsc_chi_boundary}
\end{equation}
The solution of Eq.~(\ref{eq:nsc_chi_equation}) regular at the origin and satisfying the boundary condition~(\ref{eq:nsc_chi_boundary}) reads as
\begin{equation}
    \chi_\omega(r) = 2e^2\nu R_s \frac{ J_1(q_\omega r)}{ q_\omega R_sJ_0(q_\omega R_s)-J_1(q_\omega R_s)} \ ,
    \label{eq:nsc_chi_solution}
\end{equation}
with $q_\omega^2= i\omega/ D$.
The resulting Eqs.~(\ref{eq:nsc_diffusive_current}), (\ref{eq:nsc_rho_decomposition}) and (\ref{eq:nsc_chi_solution}) represent the solution of the linear response in the non-self-consistent approximation.

We now use the first-order solution to construct the dc component of the second-order response. The second-order correction for the distribution function is
defined by
\begin{equation}
    \label{eq:nsc_f2_expansion}
    f^{(2)}(\mathbf r,\mathbf p,t) = f_{\rm dc}^{(2)}(\mathbf r,\mathbf p) + \left[f_{2\omega}^{(2)}(\mathbf r,\mathbf p)e^{-2i\omega t} + {\rm c. c.}\right] \ .
\end{equation} 
In the same diffusive reduction
\begin{equation}
    f_{\rm dc}^{(2)}(\mathbf{r},\mathbf{p}) = f_{{\rm dc},0}^{(2)}(\mathbf{r},\xi) + \hat{\mathbf p}\cdot \mathbf f_{{\rm dc},1}^{(2)}(\mathbf{r},\xi) \ ,
    \label{eq:nsc_f2_truncation}
\end{equation}
one obtains the following expression for the dc current (details of the derivation are provided in Appendix~\ref{app:diffusive_reduction}):
\begin{equation}
    \mathbf j_{\rm dc}^{(2)}  = -D\nabla_{\mathbf r}\delta\rho_{\rm dc}^{(2)} + \frac{2e\tau}{m^*} \rm{Re} \left[\mathbf{E}_\omega^{{\rm ext}*}\delta\rho_\omega^{(1)}\right] \ .
    \label{eq:nsc_second_order_current_vector}
\end{equation}
The first term on the right-hand side of this equation is a gradient contribution. It does not contribute to the angular-independent azimuthal current. The gradient of the angular-independent ($m = 0$) part of $\delta \rho^{(2)}_{\rm dc}$ is purely radial, whereas the $m = \pm 2$ components ($\propto \exp(\pm 2i\varphi)$) remain angular dependent and therefore do not enter $j_{{\rm dc}, \varphi}^{(2)}(r)$. Thus, the inverse Faraday effect in the non-self-consistent diffusive theory is determined solely by the bilinear term in Eq.~(\ref{eq:nsc_second_order_current_vector}). Using the first-order solution~\eqref{eq:nsc_rho_decomposition} and (\ref{eq:nsc_chi_solution}), one finds
\begin{eqnarray}
    \label{eq:nsc_jphi_final}
    j_{\varphi}^{(2)}(r) = \frac{4e^3\nu\tau}{m^*} E_0^2 P_{\rm circ}\\
    \nonumber
    \times{\rm Im}\left[\frac{R_sJ_1(q_\omega r)}{q_\omega R_sJ_0(q_\omega R_s)-J_1(q_\omega R_s)}\right] \ ,    
\end{eqnarray}
where $P_{\rm circ} = (|E_-|^2-|E_+|^2)/E_0^2$ is the degree of circular polarization and $E_0^2 = |E_-|^2 + |E_+|^2$. Thus, the dc current changes sign when the radiation helicity is reversed and vanishes for linearly polarized radiation. The magnetic moment follows from Eq.~\eqref{eq:Mz_def}. It reads as
\begin{eqnarray}
   \label{eq:nsc_Mz_final}
    M_{z} = \frac{4e^3\nu\tau S}{m^*c} E_0^2 P_{\rm circ} \\
    \nonumber
    \times {\rm Im}\left[\frac{R_sJ_2(q_\omega R_s)}{q_\omega\left[q_\omega R_sJ_0(q_\omega R_s)-J_1(q_\omega R_s)\right]}\right] \ .
\end{eqnarray}
Here $S=\pi R_s^2$ is the disk area. The limiting behavior for $M_z$ in the above expression follows from the asymptotic expansion of the Bessel functions. Expanding Eq.~\eqref{eq:nsc_jphi_final} in the limit $|q_\omega| R_s\ll1$, one obtains 
\begin{equation}
    j_{\varphi }^{(2)}(r) \simeq \frac{e^3\nu\tau}{2m^*} \frac{\omega}{D}E_0^2P_{\rm circ} r \left(3R_s^2-r^2\right),
    \label{eq:nsc_jphi_low}
\end{equation}
and therefore for $|q_{\omega}|R_s \ll 1$
\begin{equation}
    \label{eq:nsc_Mz_low}
    M_{z} \simeq \frac{7\pi e^3\nu\tau}{24m^*c} \frac{\omega R_s^6}{D}E_0^2P_{\rm circ} \ .
\end{equation}
Thus, in the non-self-consistent approximation the magnitude of the response grows linearly with frequency. In the opposite regime $|q_\omega| R_s \gg 1$, while still keeping $\omega\tau\ll1$, the magnetic moment decreases as
\begin{equation}
    M_{z}\simeq\frac{S}{c}\frac{e^3v_F^2\tau^2}{\pi\omega}E_0^2P_{\rm circ} \ .
    \label{eq:nsc_Mz_high}
\end{equation}
Equations~\eqref{eq:nsc_Mz_low} and \eqref{eq:nsc_Mz_high} imply that the non-self-consistent $|M_z(\omega)|$ dependence has a maximum at the crossover between the low- and high-frequency regimes. This crossover occurs at a radiation frequency determined by the condition $\omega \sim D/R_s^2$.

\begin{figure}[!htpb]
\centering
\includegraphics[scale=0.8]{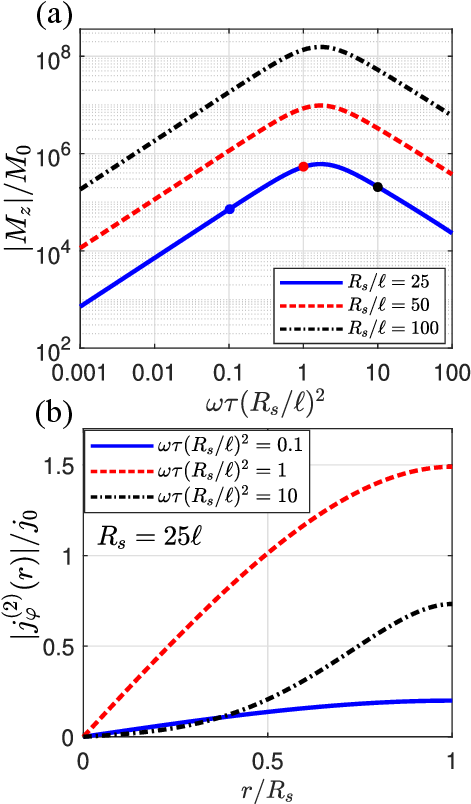}
\caption{Results of the non-self-consistent approximation. (a) Typical frequency dependencies of the induced magnetic moment~(\ref{eq:nsc_Mz_final}) for $R_s/\ell = 25$, 50, and 100. Here $M_0 = |e|^3\nu \tau E_0^2\ell^4/m^*c$. (b) Radial profiles of the second-order dc circulating current density~(\ref{eq:nsc_jphi_final}). Here $j_0 = |e|^3\nu\tau R_sE_0^2/m^*$, $R_s/\ell = 25$, and we choose the radiation frequencies $\omega\tau (R_s/\ell)^2 = 0.1$, 1, 10 (highlighted by dots in (a)). }
\label{non_self_consistent_figure}
\end{figure}

Typical results for the radial profiles of the second-order dc circulating current~(\ref{eq:nsc_jphi_final}) and the induced magnetic moment~(\ref{eq:nsc_Mz_final}) are shown in Fig.~\ref{non_self_consistent_figure}. The behavior of the second-order dc response can be qualitatively explained as follows. In the non-self-consistent problem, the only relaxation mechanism for the boundary-induced charge modulation is diffusion. At low frequencies $\omega \ll D/R_s^2$, diffusion is fast on the time scale of the ac drive. The boundary-induced charge distribution is therefore quasistatic to leading order, and this leading contribution does not generate the circulating dc current. The inverse Faraday response is produced only by the first dynamic correction and is consequently small. At frequencies of order $\omega \sim D/R_s^2$, the drive period becomes comparable to the diffusion time across the disk, and the finite-size diffusive mode gives the largest contribution to the dc circulating current. At higher frequencies, $\omega \gg D/R_s^2$, the diffusion length during one period becomes shorter than the sample radius. The charge modulation can no longer form a sample-wide profile, and the radial structure of the response leads to a suppression of the integrated current and magnetic moment.

\section{Self-consistent kinetic formulation}
\label{sec:self_consistent_response_section}

In the non-self-consistent diffusive problem, the hierarchy of distribution-function harmonics can be closed at the isotropic and first angular harmonics. This gives the diffusion equation used in the previous section and corresponds to the $|n|\leq 1$ truncation. However, in the self-consistent problem the induced electric field is a nonlocal functional of the charge density [see Eqs.~(\ref{eq:phi_m_hankel}) and~(\ref{eq:Eind_general_m})]. As a result, the charge density and the induced field can develop sharper radial features, especially near the sample boundary. For this reason, we also solve the self-consistent Coulomb problem within the $|n|\leq 2$ truncation and compare the results with those obtained in the $|n|\leq 1$ approximation. Since the kinetic-equation approach requires a boundary condition for the distribution function rather than only for the current density, we impose specular reflection at the disk edge. This condition enforces zero normal flux and corresponds to a clean boundary, where electrons are reflected without additional diffuse scattering or boundary relaxation. Therefore, the comparison between the self-consistent and non-self-consistent results directly probes the effect of Coulomb screening, while the comparison between the two harmonic truncations shows whether the usual diffusive closure remains sufficient for the self-consistent problem.

Below we provide the main technical details of the self-consistent approach. Readers not interested in these details may proceed to Sec.~\ref{sec:numerical_results}.

\subsection{Linear response}
\label{sec:self_consistent_linear}

We now formulate the self-consistent linear problem in a form adapted to the circular geometry. The distribution function depends on the azimuthal angle $\varphi$ in coordinate space and on the direction of the quasiparticle momentum, described by the angle $\varphi_{\mathbf{p}}$, in momentum space. For calculations it is particularly useful to make the change of variables and replace $\varphi_{\mathbf{p}}$ by the angle measured with respect to the local radial direction,
\begin{equation}
    \theta=\varphi_{\mathbf{p}} - \varphi \  .
    \label{eq:theta_definition}
\end{equation}
The main advantage of this step is that for specular reflection the boundary condition at the disk edge becomes independent of the position along the boundary. Indeed, in terms of the laboratory momentum angle, reflection from
the circular boundary corresponds to the shift $\varphi_{\mathbf{p}} \to 2\varphi+\pi-\varphi_{\mathbf{p}}$, whereas in terms of $\theta$ it simply maps  $\theta\to \pi-\theta$. As a result, the reflection rule is independent of the position $\varphi$
along the boundary, and the boundary condition takes the form 
\begin{equation}
    f(R_s,\varphi,\theta,p) = f(R_s,\varphi,\pi-\theta,p) \ ,
    \label{eq:specular_theta}
\end{equation}
which leads to simple algebraic relations between angular harmonics written below.

Due to the fact that the external field contains only the angular harmonics $e^{\pm i\varphi}$ [see, e.g., Eqs.~(\ref{helicity_decompositions})], the linear response splits into two independent angular sectors. Thus, considering Eq.~(\ref{eq:boltzmann}), we use the convention~Eq.~(\ref{eq:nsc_f1_expansion}) and seek the solution in the form 
\begin{equation}
    f_{\omega}^{(1)}(r,\varphi,\theta,p) = - (\partial_{\xi} f_0) \sum_{n}\sum_{m = \pm 1} u_n^{(m)}(r)e^{im \varphi + in\theta} \ .
    \label{eq:sc_linear_expansion}
\end{equation}
Here the summation is carried out over the distribution-function harmonics $n\in \mathbb{Z}$ and the index of the angular sector $m$. Note that the sector $m=+1$ is driven by the circular component $E_-$ of the external field, whereas the sector $m=-1$ is driven by $E_+$. In what follows, the perturbative order of the linear-response amplitudes is not indicated explicitly: the superscript $(m)$ labels the angular sector $e^{im\varphi}$, not the order of perturbation theory.

The usefulness of the variable $\theta$ can also be seen directly from the operator $\mathbf{v}\cdot\nabla_{\mathbf{r}}$ in the Boltzmann equation. In polar coordinates
\begin{equation}
    \mathbf v\cdot\nabla_{\mathbf r} = v\cos\theta\,\frac{\partial}{\partial r} + \frac{v\sin\theta}{r} \left(\frac{\partial}{\partial\varphi} - \frac{\partial}{\partial\theta} \right) \ .
    \label{eq:streaming_theta}
\end{equation}
Acting on $e^{im\varphi + in\theta }$, this operator couples only neighboring harmonics, $n\to n\pm1$. Substitution of Eq.~(\ref{eq:sc_linear_expansion}) into Eq.~(\ref{eq:boltzmann}), linearized with respect to the ac field amplitude, gives the system of equations describing \textit{the full hierarchy} of the distribution function harmonics
\begin{equation}
\begin{aligned}
    \gamma_n u_n^{(m)} + \frac{v_F}{2}\bigg[\frac{d u_{n-1}^{(m)}}{dr}+\frac{1+m-n}{r}u_{n-1}^{(m)}
    \\
    + \frac{d u_{n+1}^{(m)}}{dr} + \frac{n+1-m}{r}u_{n+1}^{(m)} \bigg] = S_n^{(m)} \ ,
\end{aligned}
\label{eq:sc_linear_hierarchy}
\end{equation}
where
\begin{equation}
    \gamma_n = \begin{cases} -i\omega, & n=0 \ ,\\
        -i\omega+\tau^{-1}, & n\neq 0 \ .
    \end{cases}
    \label{eq:sc_gamma}
\end{equation}
Thus, the collision integral relaxes all anisotropic harmonics, whereas the operator $\mathbf{v}\cdot\nabla_{\mathbf{r}}$ couples neighboring angular harmonics. In particular, it couples the isotropic charge mode ($n = 0$) to the current harmonics ($n = \pm 1$). The source terms $S_n^{(m)}$ in Eq.~(\ref{eq:sc_linear_hierarchy}) are nonzero only for $n=\pm1$, because the electric field acts on the equilibrium distribution. They have the form
\begin{subequations}
\label{eq:sc_sources_linear}
\begin{align}
    S_1^{(m)} = \frac{ev_F}{2} \left[ E_{\omega, r}^{(m)}(r)-iE_{\omega, \varphi}^{(m)}(r) \right] \ ,\\
    S_{-1}^{(m)}=\frac{ev_F}{2}\left[E_{\omega, r}^{(m)}(r)+iE_{\omega, \varphi}^{(m)}(r)\right] \ .
    \end{align}
\end{subequations}
Here $E_{\omega, r}^{(m)}$ and $E_{\omega,\varphi}^{(m)}$ are the radial and azimuthal components of the total first-order field. Substitution of Eq.~(\ref{eq:sc_linear_expansion}) into the specular-reflection condition~(\ref{eq:specular_theta}) gives
the boundary condition
\begin{equation}
u_n^{(m)}(R_s) = (-1)^nu_{-n}^{(m)}(R_s) \ .
\end{equation}

The first-order observables are obtained directly from the harmonics $u_n^{(m)}$. Using Eqs.~(\ref{eq:sc_linear_expansion}) and~(\ref{charge_current_definitions}), the induced charge can be written in the following form:
\begin{equation}
\label{delta_rho_via_harmonic} 
\delta\rho(\mathbf{r},t) = \sum_{m=\pm1} 2e\nu u_0^{(m)}(r)e^{im\varphi - i\omega t} + {\rm c. c.} \ .
\end{equation}
The current density reads as
\begin{equation}
    \mathbf j (\mathbf{r}, t)  = \sum_{m=\pm1} \left[j_{r}^{(m)}(r)\mathbf e_r + j_{\varphi }^{(m)}(r)\mathbf e_\varphi \right]e^{im\varphi - i\omega t} + {\rm c. c.} \ ,
\end{equation}
with
\begin{subequations}
\begin{align}
    j_{r}^{(m)} = e\nu v_F \left( u_1^{(m)}+u_{-1}^{(m)} \right) \ ,\\
    j_{\varphi}^{(m)} = i e\nu v_F \left( u_1^{(m)}-u_{-1}^{(m)} \right) \ .
\end{align}
\end{subequations}

The total electric field entering the source terms in Eqs.~(\ref{eq:sc_sources_linear}) is obtained by summing the external field and the induced electrostatic field generated by the charge modulation in Eq.~(\ref{delta_rho_via_harmonic}). Within the first-order perturbation theory the induced field is determined from the solution of the electrostatic problem~(\ref{eq:phi_m_hankel}) and~(\ref{eq:Eind_general_m}) for $m = \pm 1$. It is convenient to introduce two integral functionals of the isotropic harmonic,
\begin{subequations}
\begin{align}
    \label{eq:I0_linear}
    I_0^{(m)}(r) = \int_0^\infty d\kappa \int_0^{R_s}dr'\,r' \kappa u_0^{(m)}(r') F(\kappa) J_1(\kappa r')J_0(\kappa r) \ ,\\
    \label{eq:I2_linear}
    I_2^{(m)}(r) = \int_0^\infty d\kappa \int_0^{R_s}dr'\,r' \kappa u_0^{(m)}(r') F(\kappa) J_1(\kappa r')J_2(\kappa r) \ .
\end{align}
\end{subequations}
Here for $\epsilon_{\rm env}=1$, $q_{\rm TF}=4\pi e^2\nu$ is the two-dimensional Thomas--Fermi screening wavenumber~\cite{Stern1967}. Restoring the dielectric constant gives $q_{\rm TF} = 4\pi e^2\nu/\epsilon_{\rm env} = 2/a_B^*$, where $a_B^*$ is the effective Bohr radius. The radial and azimuthal projections of the
total electric field in the two angular sectors are then
\begin{subequations}
\label{total_field_Eplus_linear}
\begin{align}
    \label{eq:Er_plus_linear}
    E_{\omega, r}^{(+)}(r) = E_- + \frac{q_{\rm TF}}{2e} \left[ I_2^{(+)}(r)-I_0^{(+)}(r) \right] \ ,\\
    E_{\omega, \varphi}^{(+)}(r) = iE_- - i\frac{q_{\rm TF}}{2e}\left[I_0^{(+)}(r)+I_2^{(+)}(r) \right] \ ,
\label{eq:Ephi_plus_linear}
\end{align}
\end{subequations}
for the $m = +1$ sector (proportional to $e^{i\varphi}$), and
\begin{subequations}
\label{total_field_Eminus_linear}
\begin{align}
   \label{eq:Er_minus_linear}
    E_{\omega, r}^{(-)}(r) = E_+ + \frac{q_{\rm TF}}{2e} \left[I_2^{(-)}(r)-I_0^{(-)}(r) \right] \ ,\\
    \label{eq:Ephi_minus_linear}
    E_{\omega, \varphi}^{(-)}(r)=-iE_+ + i\frac{q_{\rm TF}}{2e}\left[I_0^{(-)}(r)+I_2^{(-)}(r)\right] \ ,
\end{align}
\end{subequations}
for the $m = -1$ sector (proportional to $e^{-i\varphi}$). Substituting Eqs.~(\ref{total_field_Eplus_linear}) and (\ref{total_field_Eminus_linear}) into Eq.~(\ref{eq:sc_sources_linear}) gives the closed self-consistent linear hierarchy: the isotropic harmonic of the distribution function $u_0^{(m)}$ generates the induced field, while the induced field enters back into the source terms $S_{\pm1}^{(m)}$.

Equations~(\ref{eq:sc_linear_hierarchy})-(\ref{total_field_Eminus_linear}) form the basis for numerical solution of the self-consistent linear response problem. In numerical calculations we solve Eqs.~\eqref{eq:sc_linear_hierarchy} with two truncations. The $|n|\leq1$ truncation corresponds to the strict diffusive closure, whereas the $|n|\leq2$ truncation retains the leading higher-harmonic correction to the relation between the charge density, current, and electric field. The detailed construction of the regular variables, the Coulomb matrices, and the Chebyshev collocation scheme is provided in Appendices~\ref{app:coulomb_sampling} and \ref{app:linear_solver}.

\subsection{Self-consistent approach for the inverse Faraday effect}
\label{sec:self_consistent_ife}

We proceed to the development of the self-consistent approach for the inverse Faraday effect using the self-consistent linear solution in Sec.~\ref{sec:self_consistent_linear}. Below we briefly outline the main points of this section. First, we present the dc part of the second-order Boltzmann equation together with the self-consistent closure. Second, we discuss the expressions for the second-order and bilinear source terms in the Boltzmann equation and analyze their structure. It is demonstrated that $|n|\leq n_{\rm max}$ truncation of the distribution function in the linear problem requires $|n|\leq n_{\rm max} + 1$ truncation in the second-order dc response. Third, we provide the system of equations describing the hierarchy of the angular harmonics of the dc part of the distribution function and show that the second-order electrostatic problem does not need to be solved to determine the angular-independent circulating current. The second-order kinetic equations still have to be solved, but their source terms are fully determined by the self-consistent linear solution.

The dc part of the second-order Boltzmann equation~(\ref{eq:boltzmann}) can be written as
\begin{equation}
    \mathbf v\cdot\nabla_{\mathbf r} f_{\rm dc}^{(2)} + B(\mathbf r,\mathbf p) + Q(\mathbf r,\mathbf p) = -\frac{f_{\rm dc}^{(2)} -\langle f_{\rm dc}^{(2)}\rangle_{\mathbf p} }{\tau} \ .
    \label{eq:dc_boltzmann_sc}
\end{equation}
Here
\begin{eqnarray}
    \label{eq:dc_bilinear_source}
    B(r,\varphi,\theta,p) = 2{\rm Re}\left[e\mathbf E_\omega^{(1)*}\cdot\nabla_{\mathbf p}f_\omega^{(1)}\right] = \\
    \nonumber
     = \sum_n\sum_{m = 0,\pm2}B_{n}^{(m)}(r, p)e^{im\varphi + in\theta} \ ,
\end{eqnarray}
is the bilinear source constructed from the self-consistent first-order solution. The electric field $\mathbf E_\omega^{(1)}$ in Eq.~(\ref{eq:dc_bilinear_source}) is the \textit{total first-order field} defined in Eqs.~(\ref{total_field_Eplus_linear})--(\ref{total_field_Eminus_linear}). The second source,
\begin{eqnarray}
    \label{eq:dc_quadratic_source}
    Q(r,\varphi,\theta,p) = e\mathbf E_{\rm dc}^{(2)}\cdot\nabla_{\mathbf p}f_0 = \\
    \nonumber
    = \sum_n\sum_{m = 0,\pm 2}Q_n^{(m)}(r,p)e^{im\varphi + in\theta} \ ,
\end{eqnarray}
contains the induced second-order dc electric field. Due to the fact that the external electric field is given by linear combinations of the terms with angular structure $e^{\pm i\varphi}$ [see, e.g., Eq.~(\ref{helicity_decompositions})], the dc second-order response should contain the angular harmonics $m=0,\pm2$.  The corresponding second-order charge density and electrostatic potential can be decomposed as
\begin{subequations}
\begin{align}
    \label{eq:dc_charge_decomposition}
    \delta\rho_{\rm dc}^{(2)}(r,\varphi) = \delta \rho_0^{(2)}(r) + \delta\rho_2^{(2)}(r)e^{2i\varphi} + \delta\rho_{-2}^{(2)}(r)e^{-2i\varphi} \ ,\\
    \label{eq:dc_potential_decomposition}
    \Phi_{\rm dc}^{(2)}(r,\varphi) = \Phi_0^{(2)}(r) + \Phi_2^{(2)}(r)e^{2i\varphi} + \Phi_{-2}^{(2)}(r)e^{-2i\varphi} \ .
\end{align}
\end{subequations}
The second-order electrostatic problem is solved using the Coulomb closure
introduced in Sec.~\ref{sec:basic_equations}. In the remainder of this section, all second-order electric fields and
potentials refer to the dc component. We omit the subscript dc in their angular components for brevity. Using Eq.~(\ref{eq:phi_m_hankel}) with the second-order induced charge density $\delta\rho_m^{(2)}$, one obtains the second-order correction for the electrostatic potential
\begin{eqnarray}
    \label{eq:dc_poisson_closure}
    \Phi_m^{(2)}(r) = 2\pi \int_0^\infty d\kappa \int_0^{R_s}dr'\,r' F(\kappa)\\
    \nonumber
    \times J_m(\kappa r)J_m(\kappa r')
    \delta\rho_m^{(2)}(r') \ , \ \ \ 
    m=0,\pm2 \ ,    
\end{eqnarray}
and the projections of the second-order induced electric field are as follows: 
\begin{equation}
    E_{r,m}^{(2)}(r) = -\frac{d\Phi_m^{(2)}}{dr} \ , \ \ E_{\varphi,m}^{(2)}(r) = -\frac{im}{r}\Phi_m^{(2)}(r) \ .
    \label{eq:dc_field_components}
\end{equation}
Thus, the general theory of the second-order dc response requires taking into account the induced second-order dc electric field, which should be determined self-consistently with the second-order charge density modulation.

Let us now analyze the structure of the source terms. The angular structure of the coefficients $Q_n^{(m)}$ defined by Eq.~(\ref{eq:dc_quadratic_source}) is rather simple and fixed by the fact that the equilibrium distribution is isotropic in momentum space. Indeed,
\begin{equation}
    \nabla_{\mathbf p} f_0 = \hat{\mathbf p}\,\partial_p f_0 = \left( \mathbf e_r \cos\theta + \mathbf e_\varphi \sin\theta \right)\partial_p f_0 \ ,
    \label{eq:gradp_f0_theta}
\end{equation}
and therefore for a given spatial harmonic $m$,
\begin{equation}
    e\mathbf{E}_{\rm dc}^{(2)}\cdot\nabla_{\mathbf{p}} f_0 = e \left[E^{(2)}_{r,m}\cos\theta + E^{(2)}_{\varphi,m}\sin\theta \right]\partial_p f_0 \ .
    \label{eq:Qm_theta}
\end{equation}
It can be seen from the above equation that the coefficients $Q_n^{(m)}$ are nonzero only for $n=\pm1$:
\begin{subequations}
 \label{eq:Q_pm_explicit}
\begin{align}
    Q_1^{(m)}(r,p) = \frac{e}{2} \left( E_{r,m}^{(2)}-iE_{\varphi,m}^{(2)} \right) \partial_p f_0 \ ,\\
    Q_{-1}^{(m)}(r,p) = \frac{e}{2}\left(E_{r,m}^{(2)}+iE_{\varphi,m}^{(2)}\right)\partial_p f_0 \  .
\end{align}
\end{subequations}
We now consider the bilinear source~(\ref{eq:dc_bilinear_source}) and identify which of its components are relevant for the inverse Faraday response. Using the angular decomposition of the self-consistent linear solution~(\ref{eq:sc_linear_expansion}), which can be rewritten as 
\begin{subequations}
\begin{align}
    f_\omega^{(1)}(r,\varphi,\theta,p) = \sum_{s=\pm1} e^{is\varphi}F^{(s)}(r,\theta,p) \ ,\\
    \label{eq:F_sigma_def}
    F^{(s)}(r,\theta,p) = -(\partial_\xi f_0) \sum_n u_n^{(s)}(r)e^{in\theta} \ ,
\end{align}
\end{subequations}
together with the analogous expansion of the total first-order electric field
\begin{equation}
    \mathbf E_\omega^{(1)} = \sum_{s=\pm1} e^{is\varphi} \left[ E_{\omega, r}^{(s)}(r)\mathbf e_r + E_{\omega, \varphi}^{(s)}(r)\mathbf e_\varphi \right] \ ,
    \label{eq:E_sigma_decomp}
\end{equation}
one can see that the bilinear source~(\ref{eq:dc_bilinear_source}) contains products of the form $e^{-is\varphi}e^{is'\varphi}$. The products with $s'=s$ generate the angular-independent component $m=0$, whereas the products with $s'\neq s$ generate the components $m=\pm2$. Due to the fact that the orbital inverse Faraday effect is determined by the angular-independent circulating current, only the $m=0$ part of $B$ is needed for calculations. Thus, for the angular-independent sector we can write
\begin{eqnarray}
    \label{eq:B0_decomp}
    B^{(0)}(r,\theta,p) = \sum_n B_n^{(0)}(r,p)e^{in\theta} = \\
    \nonumber 
    = 2{\rm Re} \sum_{s=\pm1} {\mathcal B}_s(r,\theta,p) \ ,    
\end{eqnarray}
where
\begin{eqnarray}
\label{eq:B_sigma_structure}
    {\mathcal B}_s(r,\theta,p) = e\bigg\{\left[ E_{\omega, r}^{(s)*}\cos\theta + E_{\omega, \varphi}^{(s)*}\sin\theta \right]\partial_p F^{(s)} \ \ \  \\
    \nonumber 
    + \frac{1}{p} \left[ -E_{\omega, r}^{(s)*}\sin\theta + E_{\omega, \varphi}^{(s)*}\cos\theta \right]\partial_\theta F^{(s)} \bigg\} \ .
\end{eqnarray}
The resulting Eq.~(\ref{eq:B_sigma_structure}) illustrates that the bilinear sources entering the $m=0$ sector are determined solely by the self-consistent first-order harmonics $u_n^{(s)}$ and by the total first-order fields $E_{\omega, r}^{(s)}$, $E_{\omega, \varphi}^{(s)}$. It also reveals the harmonic content of the bilinear source in the momentum space. Indeed, since the factors $\cos\theta$ and $\sin\theta$ shift the momentum harmonic by one, for the truncation $|n|\leq n_{\rm max}$ in the self-consistent linear problem the bilinear source $B^{(0)}$ contains momentum harmonics up to $|n|=n_{\rm max} + 1$.

We now show that the second-order electrostatic problem does not need to be solved to determine the angular-independent circulating current. The latter is still obtained from the second-order kinetic equations, but the corresponding source terms are fully determined by the self-consistent linear solution.
For this purpose, we first introduce the expansion of the second-order dc correction for the distribution function
\begin{equation}
    f_{\rm dc}^{(2)}(r,\varphi,\theta,p) =  \sum_{m=0,\pm2}  e^{im\varphi} \sum_n G_n^{(m)}(r,p)e^{in\theta} \ .
    \label{eq:dc_expansion}
\end{equation} 
Substitution of Eq.~(\ref{eq:dc_expansion}) into Eq.~(\ref{eq:dc_boltzmann_sc}) gives the hierarchy
\begin{eqnarray}
\label{eq:dc_hierarchy}
    \Gamma_n G_n^{(m)} + \frac{v}{2} \bigg[\left(\frac{\partial}{\partial r} + \frac{m-n+1}{r} \right) G_{n-1}^{(m)} \ \ \\
    \nonumber
    + \left( \frac{\partial}{\partial r} + \frac{n+1-m}{r} \right) G_{n+1}^{(m)} \bigg] = -B_n^{(m)}-Q_n^{(m)}  \ ,
\end{eqnarray}
where
\begin{equation}
    \Gamma_n =
    \begin{cases}
        0, & n=0 \ ,\\
        \tau^{-1}, & n\neq0 \ .
    \end{cases}
    \label{eq:dc_gamma}
\end{equation}
The solution of Eq.~(\ref{eq:dc_hierarchy}) for $m = 0$ determines the azimuthal current density responsible for the induced magnetic moment. In particular, using the expansion~(\ref{eq:dc_expansion}) and the definition of the current density~(\ref{current_density_definition}), one finds 
\begin{equation}
    j^{(2)}_\varphi(r)  = ie\int\frac{pdp}{2\pi} v\left[ G^{(0)}_1(r,p)-G^{(0)}_{-1}(r,p) \right] \ .
    \label{eq:jphi_G_difference}
\end{equation}
Direct analysis of Eq.~(\ref{eq:dc_hierarchy}) reveals that the hierarchy equations for the following antisymmetric harmonics 
\begin{equation}
    A_i(r,p) = G^{(0)}_i(r,p)-G^{(0)}_{-i}(r,p) \ , \ \
    i=1,2,3 \ldots  \ ,
    \label{eq:Ai_definition}
\end{equation}
form a closed system. For instance, below we write the explicit system for $|n|\leq 2$ truncation for the self-consistent linear problem ($|n|\leq 3$ truncation for the second-order dc response)
\begin{subequations}
\label{second_order_dc_hierarchy}
\begin{align}
    \label{eq:A1_preint}
    \tau^{-1}A_1+\frac{v}{2}\left(\frac{\partial}{\partial r}+\frac{2}{r}\right)A_2= \\
    \nonumber
   = -\left(B^{(0)}_1-B^{(0)}_{-1}\right)-\left(Q^{(0)}_1-Q^{(0)}_{-1}\right) \ ,\\
    \label{eq:A2_preint}
    \tau^{-1}A_2 + \frac{v}{2} \bigg[\left(\frac{\partial}{\partial r}-\frac{1}{r}\right)A_1+\left(\frac{\partial}{\partial r}+\frac{3}{r}\right) A_3 \bigg] = \\
    \nonumber
    =  -\left(B^{(0)}_2-B^{(0)}_{-2}\right) \ ,\\
    \label{eq:A3_preint}
    \tau^{-1}A_3+\frac{v}{2}\left(\frac{\partial}{\partial r}-\frac{2}{r}\right)A_2= \\
    \nonumber
   = -\left(B^{(0)}_3-B^{(0)}_{-3}\right) \ .
    \end{align}
\end{subequations}
The second-order electric-field source only enters Eq.~(\ref{eq:A1_preint}) in combination $Q_1^{(0)}-Q_{-1}^{(0)}$. Using Eqs.~(\ref{eq:Q_pm_explicit}) and the field projections in Eq.~(\ref{eq:dc_field_components}), one sees that the $m = 0$ part of the second-order electric field is purely radial $E_{\varphi,0}^{(2)} = 0$ and
\begin{equation}
    Q^{(0)}_1 = Q^{(0)}_{-1} = \frac{e}{2}E^{(2)}_{r,0}\partial_p f_0 \ .
    \label{eq:Q_equal_m0}
\end{equation}
Thus, the antisymmetric combination $Q_1^{(0)}-Q_{-1}^{(0)} = 0$, and the  second-order electrostatic field does not enter the closed subsystem for the angular-independent circulating current. This is the key simplification: the second-order electrostatic field affects the second-order charge distribution and radial response, but it drops out of the difference of harmonics which determines the angular-independent circulating current.

Equations~(\ref{eq:dc_hierarchy})--(\ref{second_order_dc_hierarchy}) form the basis of the numerical approach for the
self-consistent calculation of the inverse Faraday effect. These equations are solved using the Chebyshev collocation method for two truncations. In particular, for the results of the self-consistent linear response for $|n|\leq 1$ ($|n|\leq 2$) truncation, the second-order dc response is calculated for $|n|\leq 2$ ($|n|\leq 3$) truncation. Details of the numerical approach are provided in Appendix~\ref{app:second_order_solver}.

\section{Numerical results and discussion}
\label{sec:numerical_results}

\subsection{Linear response}
\label{sec:numerical_results_linear_response}

\begin{figure*}[!htpb]
\centering
\includegraphics[scale=0.7]{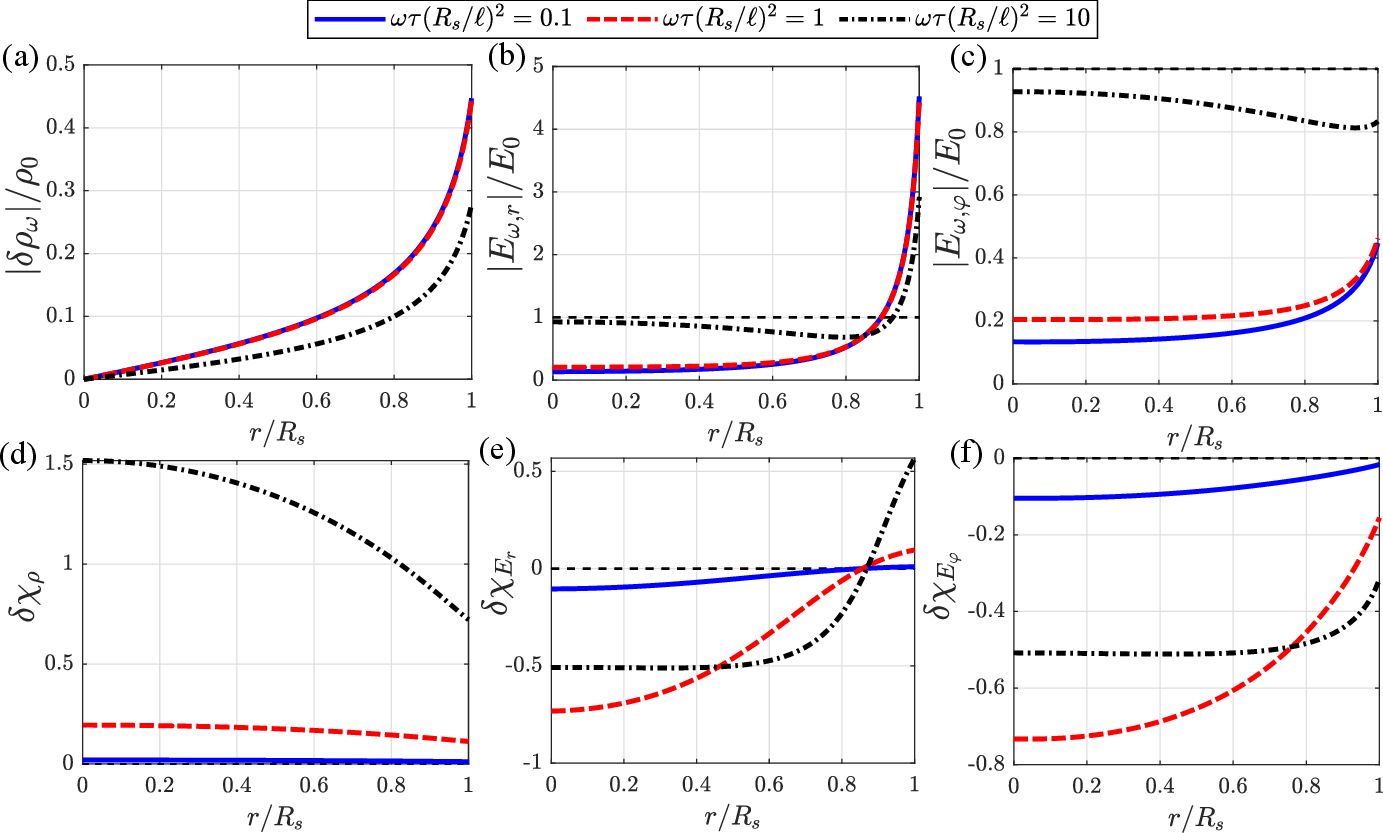}
\caption{Self-consistent linear response in the disk. (a) Radial profile of the first-order charge modulation amplitude $|\delta\rho_\omega|/\rho_0$, where $\rho_0=2e^2\nu E_0R_s$. (b), (c) Radial profiles of the ac electric-field amplitudes $|E_{\omega,r}|/E_0$ and $|E_{\omega,\varphi}|/E_0$. (d)--(f) Corresponding phase shifts $\delta\chi_\rho$,  $\delta\chi_{E_r}$, and $\delta\chi_{E_\varphi}$. The field phases are measured relative to the external-field components $E_{\omega, r}^{\rm ext}=E_0$ and $E_{\omega, \varphi}^{\rm ext}=iE_0$. The curves correspond to $\omega\tau(R_s/\ell)^2=0.1,1,10$. Dashed horizontal lines in (b),(c) and (e),(f) indicate the values for the spatially uniform external field. Parameters are $R_s/\ell=25$, $q_{\rm TF}\ell=0.4$, and $d/\ell=1$.}
\label{fig:linear_response_profiles}
\end{figure*}

We proceed to the discussion of the results of numerical calculations. We first analyze the self-consistent linear response. As shown in Sec.~\ref{sec:self_consistent_response_section}, the second-order dc source is determined by bilinear products of the first-order distribution function and the total first-order electric field. Therefore, before discussing the rectified current it is useful to understand how the ac charge modulation and the local electric field are modified by Coulomb screening. Unless stated otherwise, the numerical results in this section are evaluated using the $|n|\le 2$ truncation of the first-order kinetic hierarchy and the corresponding $|n|\le 3$ truncation of the second-order dc problem. The non-self-consistent reference curves are calculated within the same collocation scheme by setting the Coulomb coupling to zero, $\lambda=q_{\rm TF}R_s=0$, while keeping all other parameters and boundary conditions unchanged.

Typical radial profiles of the charge modulation and of the total electric field are shown in Figs.~\ref{fig:linear_response_profiles}(a)--(c). Since the external electric field is spatially homogeneous in the disk plane, it does not create a bulk charge source through $\nabla\cdot\mathbf E^{\rm ext}$. The arising charge modulation is instead a finite-size effect. The distribution function and the induced electrostatic field adjust so that the normal component of the current vanishes at the disk edge. As a result, the first-order charge density acquires a pronounced radial dependence and its amplitude increases toward the boundary, as shown in Fig.~\ref{fig:linear_response_profiles}(a).

This boundary-induced charge density generates a scalar electric potential, which acts back on the electron motion and changes the local field entering the kinetic equation. This effect is illustrated in Figs.~\ref{fig:linear_response_profiles}(b) and \ref{fig:linear_response_profiles}(c), where we show the amplitudes of the radial and azimuthal components of the total first-order field. It can be clearly seen that the radial component is enhanced near the boundary, while the azimuthal component is reduced in the bulk compared with the corresponding one of the external field. Thus, the self-consistent field is not simply the externally applied homogeneous field with a small correction. The total field becomes spatially nonuniform across the sample. At low frequencies the charge redistribution is close to quasistatic, and we observe that the screening correction is strongest. With increasing frequency the charge modulation cannot fully follow the ac drive, so the induced electrostatic field becomes less efficient and the total field moves closer to the external one.

The phase-shift profiles shown in Figs.~\ref{fig:linear_response_profiles}(d)--(f) illustrate an additional important point. The phase of the charge modulation and the phases of both components of the total electric field vary across the disk. This spatially inhomogeneous phase lag is a characteristic feature of the finite-size self-consistent response. It shows that the effect of screening cannot be represented by a single frequency-dependent scalar factor multiplying the external field. Instead, Coulomb interaction produces a complex-valued field profile with both spatially dependent amplitude and spatially dependent phase. This distinction is important for the inverse Faraday effect, because the second-order dc source contains products of first-order quantities. Consequently, the resulting circulating dc current is sensitive not only to the magnitudes of the linear-response fields, but also to their relative phases.

\begin{figure}[!t]
\centering
\includegraphics[scale=0.85]{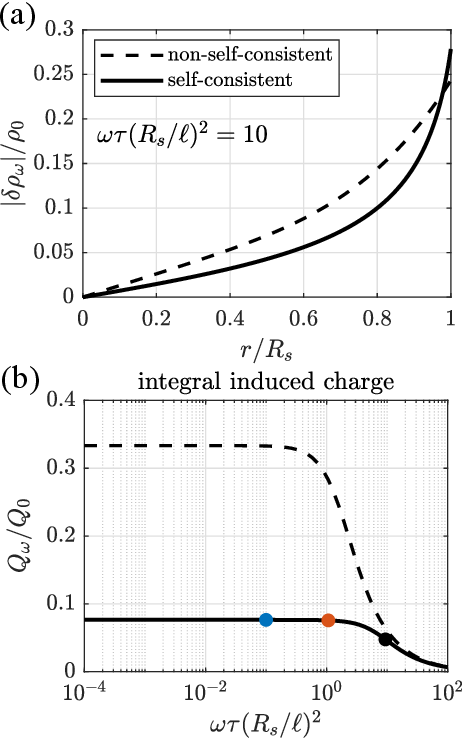}
\caption{Effect of Coulomb screening on the first-order charge modulation. (a) Comparison of the radial profiles of $|\delta\rho_\omega|/\rho_0$ obtained in the self-consistent and non-self-consistent calculations at $\omega\tau(R_s/\ell)^2=10$. (b) Frequency dependence of the integral charge modulation $Q_\omega=2\pi\int dr\,r\,|\delta\rho_\omega(r)|$, normalized by $Q_0=2q_{\rm TF}E_0R_s^3$. Colored dots mark the frequencies used in Fig.~\ref{fig:linear_response_profiles}. The non-self-consistent result is obtained by setting $\lambda=q_{\rm TF}R_s=0$ in the same numerical scheme. The parameters are the same as in Fig.~\ref{fig:linear_response_profiles}.}
\label{fig:linear_response_charge_screening}
\end{figure}

\begin{figure}[!htpb]
\centering
\includegraphics[scale=0.53]{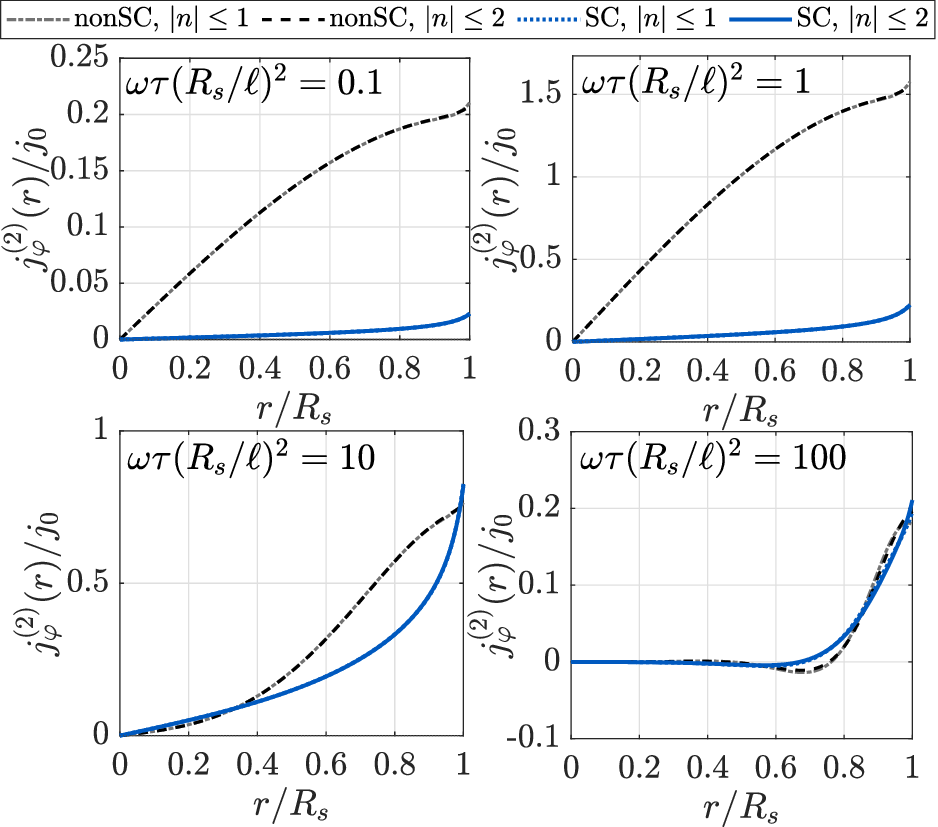}
\caption{Radial profiles of the second-order dc circulating current density $j_{\varphi}^{(2)}(r)$ for $\omega\tau(R_s/\ell)^2=0.1,1,10,100$. The self-consistent (SC) and non-self-consistent (nonSC) results are compared for the $|n|\le 1$ and $|n|\le 2$ truncations of the first-order kinetic hierarchy. The non-self-consistent curves are obtained by setting $\lambda=q_{\rm TF}R_s=0$ in the same collocation scheme, while the self-consistent curves correspond to $q_{\rm TF}\ell=0.4$. The current density is normalized to $j_0=|e|^3\nu\tau R_sE_0^2/m^*$. The profiles are shown for $P_{\rm circ}=-1$; reversing the radiation helicity changes the sign of $j_{\varphi}^{(2)}(r)$. Parameters are $R_s/\ell=25$ and $d/\ell=1$.}
\label{fig:second_order_current_profiles}
\end{figure}

\begin{figure*}[!htpb]
\centering
\includegraphics[scale=0.72]{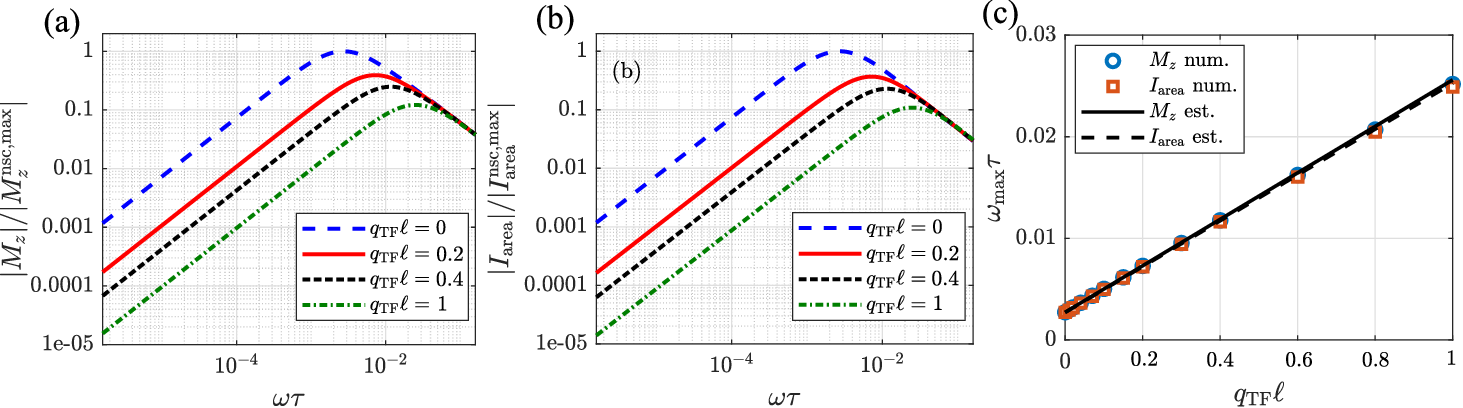}
\caption{Frequency dependence and Coulomb-induced peak shift of the second-order dc response. (a),(b) Absolute values of the induced magnetic moment $M_z$ and the area-integrated circulating current $I_{\rm area}$, normalized by their non-self-consistent peak values, for several screening strengths $q_{\rm TF}\ell$. The curve with $q_{\rm TF}\ell=0$ corresponds to the non-self-consistent result. (c) Peak position $\omega_{\rm max}\tau$ extracted from the numerical curves and compared with the geometry-corrected screened-diffusive estimate~(\ref{estimate_expression}). Parameters are $R_s/\ell=25$ and $d/\ell=1$. For this set we obtain the following dimensionless geometrical factor: $\eta_{\rm geom}\approx 0.64$ for $|M_z|$ and $\eta_{\rm geom}\approx 0.63$ for $|I_{area}|$.}
\label{fig:second_order_integral_response}
\end{figure*}

The effect of screening on the charge modulation is further clarified in Fig.~\ref{fig:linear_response_charge_screening}. Panel (a) compares the self-consistent and non-self-consistent charge profiles at $\omega\tau(R_s/\ell)^2=10$. In both calculations the charge modulation is generated by the finite boundary and increases toward the edge of the disk. However, the electrostatic feedback suppresses its magnitude in the bulk and changes the radial shape of the profile. The self-consistent charge distribution therefore cannot be obtained from the non-self-consistent one by a simple overall rescaling. Screening redistributes the charge response over the sample and modifies the boundary charge buildup.

The frequency dependence of the integrated charge modulation $Q_\omega=2\pi\int dr\,r|\delta\rho_\omega(r)|$ is shown in Fig.~\ref{fig:linear_response_charge_screening}(b). We see that in the  non-self-consistent problem, the low-frequency response approaches a quasistatic finite-size profile. Therefore $Q_\omega$ remains approximately constant at small $\omega\tau(R_s/\ell)^2$. This behavior does not contradict the low-frequency behavior of the inverse Faraday current. The latter vanishes in the quasistatic limit because the leading charge profile does not by itself generate an angular-independent dc circulating current. At higher frequencies, when the diffusion length $\sqrt{D/\omega}$ becomes shorter than the sample radius, the integrated charge modulation decreases. The self-consistent curve lies well below the non-self-consistent one  already in the low-frequency regime. This shows that Coulomb screening suppresses the boundary-induced charge accumulation even when the response is nearly quasistatic. At larger frequencies the self-consistent integrated charge decreases further, reflecting the combined effect of finite diffusive relaxation and screening. Thus, the self-consistent linear response differs from the non-self-consistent one in two important ways: the induced charge distribution is suppressed and reshaped, and the local electric field entering the nonlinear source acquires a nontrivial spatial dependence in both amplitude and phase.

\subsection{Coulomb screening of the inverse Faraday effect}
\label{sec:numerical_results_second_order}

We now turn to the second-order dc response. Before discussing the integral observables, it is useful to inspect the radial profile of the circulating current density $j_{\varphi}^{(2)}(r)$, since both $I_{\rm area}$ and $M_z$ are obtained as radial moments of this distribution. The corresponding profiles are shown in Fig.~\ref{fig:second_order_current_profiles} for several representative driving frequencies. Fig.~\ref{fig:second_order_current_profiles} also compares the $|n|\le 1$ and $|n|\le 2$ truncations of the first-order kinetic hierarchy. The two truncations give close results for both the self-consistent and non-self-consistent calculations over the shown frequency range. This indicates that the main conclusions, which are presented in this section, are not sensitive to the retained number of angular harmonics. In the following, the integral observables are evaluated using the $|n|\le 2$ truncation of the first-order problem and the corresponding $|n|\le 3$ truncation of the second-order dc problem.

The comparison between the self-consistent and non-self-consistent profiles shows how Coulomb screening modifies the second-order dc current distribution. At low frequencies [see Figs.~\ref{fig:second_order_current_profiles}(a) and \ref{fig:second_order_current_profiles}(b)] the self-consistent current is strongly suppressed relative to the non-self-consistent result. This suppression is consistent with the above discussed features of the linear response. Screening reduces the boundary-induced charge accumulation and changes the local total electric field entering the bilinear source. Around $\omega\tau(R_s/\ell)^2=10$, the self-consistent current becomes more concentrated near the disk edge and can reach values comparable to the non-self-consistent profile close to the boundary, while remaining smaller in the bulk. At higher frequencies, the current profile is confined to a narrow boundary region and the total integrated response is reduced. Thus, Coulomb screening affects both the magnitude of the second-order dc circulating current and its radial profile.

The consequences of this redistribution for integral observables are shown in Fig.~\ref{fig:second_order_integral_response}. Panels (a) and (b) present the frequency dependence of the induced magnetic moment $M_z$ and the area-integrated circulating current $I_{\rm area}$. In the non-self-consistent limit, the response has a maximum near the bare finite-size diffusive scale $\omega\sim D/R_s^2$. One can clearly see that the increase in the Coulomb coupling suppresses the peak amplitude and shifts the maximum of integral observables to higher frequencies. The fact that both $M_z$ and $I_{\rm area}$ show the same trend indicates that the effect is not an artifact of the particular radial weighting used to define the integral observables. At the same time, the two observables need not be identical, because $M_z$ contains an additional power of $r$ and is therefore more sensitive to currents near the sample edge.

To understand the observed shift of the maximum of $M_z$, it is useful to relate the numerical result to the screened relaxation rate of the \textit{first-order} charge modulation. The simplest way to obtain this estimate is to consider the infinite-plane limit, within which the Hankel transform diagonalizes the nonlocal Coulomb operator. The finite disk is then described by replacing the wave number $q$ by an effective wave number $q_*$ associated with the dominant finite-size charge mode.

Without Coulomb interaction, a charge mode $\delta\rho_{\mathbf{q}}$ with wave number $q$ relaxes diffusively. Indeed, the continuity equation together with the diffusive current gives $\partial_t \delta\rho_{\mathbf{q}}^{(1)} = -Dq^2\delta\rho_{\mathbf{q}}^{(1)}$. Thus, the bare relaxation rate is $\Gamma_0(q)=Dq^2$. In the self-consistent problem, the same charge modulation also generates the electric field. For a two-dimensional charge distribution embedded 
in three-dimensional space, the Fourier component of the potential is
\begin{equation}
\Phi_{\mathbf{q}}^{(1)} = (2\pi/q)F(qd)\delta\rho_{\mathbf{q}}^{(1)} \ .
\end{equation}
The induced electric field is $\mathbf E_q^{\rm ind}=-i\mathbf q\Phi_q$, 
and it produces an additional drift contribution to the current,
\begin{equation}
\mathbf{j}_{\mathbf{q}}^{(1)}=-i\mathbf q D\delta\rho_{\mathbf{q}}^{(1)}+\sigma \mathbf E_{\mathbf{q}}^{(1) {\rm ind}} \ .
\end{equation}
Substitution into the continuity equation gives
\begin{equation}\label{continuity_equation}
\partial_t\delta\rho_{\mathbf{q}}^{(1)} = -\left[Dq^2+2\pi\sigma qF(qd)\right]\delta\rho_{\mathbf{q}}^{(1)} \ .
\end{equation}
Using $\sigma=2e^2\nu D$ and $q_{\rm TF}=4\pi e^2\nu$, one obtains the screened relaxation rate
\begin{equation}
\label{screened_relaxation_rate}
\Gamma_{\rm scr}(q) = Dq^2+Dq_{\rm TF}qF(qd) \ .
\end{equation}
The first term is the ordinary diffusive relaxation rate, while the second term is the Coulomb contribution. Its linear dependence on $q$ reflects the $1/q$ form of the two-dimensional Coulomb potential in momentum space. For a finite disk, the response is not determined by a single plane-wave mode. Nevertheless, the peak position can be estimated by introducing an 
effective wave number $q_*$ extracted from the non-self-consistent maximum ($\omega_{\rm max}^{(0)}$): $q_*=\sqrt{\omega_{\rm max}^{(0)}/D}$. Substituting $q=q_*$ into the screened relaxation rate~(\ref{screened_relaxation_rate}) 
then gives the estimate
\begin{equation}
\label{estimate_expression}
\omega_{\rm est} = \omega_{\rm max}^{(0)} + \eta_{\rm geom}Dq_{\rm TF}q_*F(q_*d) \ .
\end{equation}
Here we introduced a dimensionless geometry factor $\eta_{\rm geom}$, which accounts for the finite-disk radial profile of the charge mode and for the projection of the nonlocal Coulomb kernel onto this profile. In the numerical calculations $\eta_{\rm geom}$ is determined from the weak-screening slope of the peak shift. A rather simple estimate of the shift in the peak position can be obtained from the above equation for $q_* \sim 1/R_s$, which yields $\delta \omega \sim Dq_{\rm TF}/R_s$.

The comparison in Fig.~\ref{fig:second_order_integral_response}(c) shows that this geometry-corrected screened-diffusive estimate agrees very well with the peak positions extracted from the full numerical solution. This agreement supports the interpretation that the displacement of the maximum is controlled primarily by the Coulomb-enhanced relaxation rate of the boundary-induced charge mode.

\subsection{Discussion}
\label{sec:discussion}

The results presented in this work were obtained within several controlled approximations. First, the analysis was carried out in the quasiclassical diffusive regime. The corresponding applicability conditions are $p_F^{-1} \ll \ell \ll R_s$ and $\omega\tau\ll 1$ together with the nonretarded condition discussed in Sec.~\ref{sec:basic_equations}. In this regime the microscopic band structure enters only through the parameters of the kinetic theory, such as the Fermi velocity, density of states, diffusion coefficient, and elastic scattering time. The mechanism considered here is therefore generic for finite-size diffusive conductors. Additional contributions to the IFE response can appear beyond the quasiclassical limit. They can originate from quantum-interference corrections, finite-level-spacing effects, effects of nonparabolic or Dirac-like dispersion~\cite{Magarill1999,Koshelev2015,Tokman2020,Tanaka2020}. While the finite-size screening mechanism discussed in this work should remain relevant whenever the response involves boundary-induced charge redistribution, the quantitative magnitude and even the relative importance of different IFE contributions can depend on the dispersion relation of charge carriers.

The charge dynamics studied in this work is overdamped and diffusive. The considered screened diffusive mode corresponds to the overdamped limit of the same self-consistent charge dynamics that gives rise to two-dimensional plasmons. To illustrate this point, we consider Eq.~(\ref{continuity_equation}) in the Coulomb-dominated long-wavelength limit and extend the analysis beyond the low-frequency diffusive regime $\omega\tau \ll 1$ by restoring the full frequency dependence of the conductivity. Neglecting the contribution $\propto q^2$ associated with the diffusion and recovering the ac Drude conductivity $\sigma \to \sigma/(1-i\omega\tau)$, one obtains the standard equation for the poles of the density response $\omega^2 + i\omega/\tau - \omega_{\rm pl}^2(q) = 0$. Here $\omega_{\rm pl}^2(q) = 2\pi \sigma qF(qd)/\tau$ is the squared plasma frequency. Within this local Drude description the boundary between overdamped and underdamped regimes is given by $\omega_{\rm pl}(q)\tau = 1/2$. The present work is restricted to the strongly overdamped regime $\omega_{\rm pl}(q)\tau \ll 1$, while for $\omega_{\rm pl}(q)\tau \gg 1$, one recovers $\omega = \pm \omega_{\rm pl}(q)-i/2\tau$.  Substantial enhancement of the inverse Faraday response in the underdamped regime has been predicted and observed in Refs.~\cite{Potashin2020,Han2023}.

The applicability conditions of the developed treatment depend on the harmonic truncation scheme used to solve the Boltzmann equation. In particular, the strict diffusive reduction of the $|n|\leq 1$ hierarchy requires $q\ell\ll1$ and $\omega\tau\ll1$. In order for the maximum of the inverse Faraday response to remain within the same frequency range, one should also require $\Gamma_{\rm scr}(q_*)\tau\ll1$. Using $\Gamma_{\rm scr}(q)=Dq^2+\tau\omega_{\rm pl}^2(q)$, this condition reads $\Gamma_{\rm scr}(q_*)\tau = (q_*\ell)^2/2 + \omega^2_{\rm pl}(q_*)\tau^2 \ll 1$. Thus, for $\omega_{\rm pl}(q_*)\tau\ll1$, together with $q_*\ell\ll1$, the maximum of the magnetization response remains safely within the diffusive regime. In the numerical calculations we also retain higher angular harmonics and compare the $|n|\leq1$ and $|n|\leq2$ truncations. The important point is that increasing the number of retained harmonics systematically extends the kinetic description beyond the strict diffusive limit by keeping higher momentum-space harmonics generated by spatial variations of the distribution function. Nevertheless, the present work is focused on the diffusive regime, and our quantitative conclusions are restricted to the parameter range for which the results are insensitive to the harmonic truncation.

To estimate the range of screening parameters compatible with the quasiclassical diffusive regime, we express $q_{\rm TF}\ell$ through the interaction parameter $r_s=\sqrt{2}/(k_Fa_B^*)$. Using $q_{\rm TF}=2/a_B^*$, one obtains $q_{\rm TF}\ell=\sqrt{2}\,r_s k_F\ell$. Thus, the condition $k_F\ell\gg1$ requires $q_{\rm TF}\ell\gg\sqrt{2}\,r_s$. For conventional quantum wells with $r_s\sim1$, this condition excludes very small values of $q_{\rm TF}\ell$. At the same time, substantially smaller values of $r_s$ are possible in small-effective-mass two-dimensional electron systems; for example, $r_s\simeq0.17$--$0.23$ has been reported for InSb quantum wells~\cite{Lei2022}. In such systems values of $q_{\rm TF}\ell$ of order unity can remain compatible with $k_F\ell>1$.

Note that in a more general case the description of the underdamped plasmonic regime doesn't necessarily require going beyond the electrostatic approximation and considering the full electrodynamic problem~\cite{FalkoKhmel1989, KukushkinVolkov2016}. The nonretarded approximation used in this work is controlled by the condition $\sqrt{\epsilon_{\rm env}}\omega/cq \ll1$. For the sample-wide mode $q_*\sim1/R_s$ this gives  $\sqrt{\epsilon_{\rm env}}\omega R_s/c\ll1$. At frequencies corresponding to the maximum of the inverse Faraday response $\omega \sim \Gamma_{\rm scr}(q_*)$ and in the strong screening regime the onset of retardation effects occurs when $\sqrt{\epsilon_{\rm env}}\Gamma_{\rm scr}(q_*)/cq_* \approx 2\pi \sigma F(q_*d)/c \sqrt{\epsilon_{\rm env}} \sim 1$.

The calculations were performed at zero temperature. Within the quasiclassical approximation and for energy-independent elastic scattering, the leading Fermi-surface contribution is not expected to acquire a strong temperature dependence. Temperature can, however, affect the transport parameters, most importantly through the elastic and inelastic relaxation times. It can also affect non-quasiclassical corrections, for example by thermal smearing of mesoscopic or band-structure-sensitive contributions. Joule heating can also influence the measured response. The absorbed power can increase the electron temperature, thereby changing the diffusion coefficient and relaxation times. Heating can also produce thermoelectric or photothermal currents, which are even in the radiation helicity. These contributions can be separated from the inverse Faraday signal by reversing or modulating the circular polarization.

Inelastic relaxation was neglected in the present model. This approximation is justified when the inelastic relaxation length
$L_{\rm in}=\sqrt{D\tau_{\rm in}}$ is larger than the sample size or, more generally, than the length scale over which the first-order charge modulation develops. If inelastic relaxation becomes important, it is expected to reduce the finite-size enhancement of the IFE response, because the driven electronic response cannot remain coherent over the whole sample. Thus, in the regime with pronounced inelastic effects the relevant length scale can be limited not only by the diffusion length $\sqrt{D/\omega}$, but also by $L_{\rm in}$.

The circular symmetry of the disk and the circular polarization of the drive field also play an important role in the structure of the IFE response. For the ideal disk geometry, the angular-independent dc circulating current is proportional to the degree of circular polarization $P_{\rm circ}=(|E_-|^2-|E_+|^2)/E_0^2$. Thus, the response changes sign when the radiation helicity is reversed and  vanishes for linearly polarized radiation. For elliptically polarized light the same mechanism gives a reduced response proportional to $P_{\rm circ}$. In systems lacking circular symmetry, or for spatially nonuniform illumination, different angular harmonics can mix~\cite{DurnevTarasenko2020,Gunyaga2023}. As a result, the rectified current is no longer characterized only by the radial profile $j_\varphi^{(2)}(r)$ and its spatial structure becomes sensitive to the sample shape and illumination profile. Such geometries can provide additional ways to control the spatial distribution of the optically induced dc current. 
 In addition to the global sample geometry, the response can also depend on microscopic scattering at the sample boundary. The calculations above assume specular reflection at the disk edge. For a rough boundary, diffuse or quasi-diffuse scattering can modify the electron dynamics within a near-edge region of width of order $\ell$ and affect the circulating current and the induced magnetic moment~\cite{Bandurin2022,Wang2022,Afanasiev2025}. A quantitative treatment of such effects requires replacing the specular-reflection condition by an appropriate boundary condition and lies outside the scope of the present work.

Let us finally discuss the magnitude of the effect. A convenient reference estimate is provided by the non-self-consistent expression for the magnetic moment, because it sets the upper scale before the Coulomb suppression is included. For a representative moderately disordered GaAs-like two-dimensional electron system, one can take the electron concentration $n_{2D}=10^{12}$~cm$^{-2}$,  $\ell=20$~nm, $R_s=0.5$~$\mu$m. Using the effective mass $m^\ast\simeq0.067m_e$, these parameters give $k_F\simeq 2.5\times 10^8$~m$^{-1}$, $k_F\ell\simeq 5$, $v_F\simeq 4.3\times 10^5$~m/s, $\tau\simeq 4.6\times 10^{-14}$~s, $D\simeq 4.3\times 10^{-3}$~m$^2$/s. The corresponding bare finite-size diffusive scale is $D/R_s^2\simeq 1.7\times 10^{10}$~s$^{-1}$ ($\omega/2\pi\approx 2.7$~GHz), so the relevant frequencies lie in the microwave to sub-terahertz range. The material parameters used in this estimate are typical for GaAs-based two-dimensional electron systems~\cite{Vurgaftman2001_BandParameters,
Chung2022_GaAsMobility}. For a local ac field $E_0=10^5\,{\rm V/m}$, the non-self-consistent peak estimate gives $|M_z| \sim 3\times 10^5$~$\mu_B$, where $\mu_B$ is the Bohr magneton. For stronger field $E_0=10^6\,{\rm V/m}$, the same estimate increases to $|M_z|\sim  3\times 10^7$~$\mu_B$. Such field amplitudes are below or comparable to those accessible in modern strong-field terahertz experiments~\cite{Nilforoushan2022_THz100kVcm}, although heating and other nonlinear effects then have to be controlled. These values should be interpreted as non-self-consistent upper reference scales, because Coulomb screening suppresses the peak response. For GaAs-based two-dimensional systems, the static dielectric constant is $\epsilon\simeq 12.9$ and $m^\ast\simeq0.063m_e$--$0.067m_e$~\cite{Vurgaftman2001_BandParameters}. These values give an effective Bohr radius of order $10\,{\rm nm}$ and a two-dimensional Thomas-Fermi wave number $q_{\rm TF}\simeq0.2\,{\rm nm}^{-1}$. For $\ell=20$~nm, this corresponds to $q_{\rm TF}\ell\sim4$, so the Coulomb screening is expected to be substantial. For the parameter set mentioned above the characteristic Coulomb-induced frequency scale $Dq_{\rm TF}/R_s$ corresponds to $Dq_{\rm TF}/(2\pi R_s) \sim 0.3$~THz.

Thus, the provided estimates for the magnetic moment suggest that the effect can be targeted experimentally with local magnetic probes. Because the moment is produced in a mesoscopic region, NV-based magnetometry and nanoscale SQUID techniques are more appropriate than conventional bulk magnetometry~\cite{Barry2020_NVSensitivity,Schwarz2015_NanoSQUID}. The sign of the inverse-Faraday signal changes when the helicity of the radiation is reversed, whereas heating-related backgrounds are expected to be even in helicity. Therefore, periodically switching between the two circular polarizations provides a direct way to separate the inverse-Faraday signal from such backgrounds. Thus, the screened orbital inverse Faraday effect appears experimentally accessible in optimized mesoscopic conductors with local magnetic detection and helicity-modulated microwave or terahertz excitation.

\section{Summary}
\label{sec:summary}

To sum up, we developed a self-consistent kinetic theory of the orbital inverse Faraday effect in finite-size diffusive conductors. In the non-self-consistent approximation, the frequency dependence of the induced dc magnetic moment has a maximum when the diffusion length associated with the ac drive becomes comparable to the disk radius. We demonstrated that Coulomb screening substantially modifies this picture. The first-order charge modulation generates a longitudinal electric field, which changes both the local ac field and the nonlinear source of the second-order circulating current. As a result, screening suppresses the peak response and shifts its maximum to a higher frequency. The same behavior is found for both the magnetic moment and the area-integrated circulating current, demonstrating that Coulomb self-consistency is essential for quantitative predictions of helicity-dependent orbital magnetization in finite-size diffusive conductors.

The sensitivity of the investigated response to the screened finite-size charge dynamics also points to possible applications of the inverse Faraday effect for characterizing charge-relaxation and screening scales in mesoscopic conductors. The self-consistent approach developed in this work can also be extended to investigate the role of Coulomb screening in other nonlinear response phenomena including the surface photogalvanic effects and the photon drag.

The data that support the findings of this article are openly available~\cite{KopasovMelnikov2026_ZenodoData}. 

\acknowledgments
The authors acknowledge S.~V.~Sharov for stimulating discussions. The work was supported by the Russian Science Foundation (Grant No. 25-12-00042) in part of analytical studies of the second-order response, and by the Ministry of Science and Higher Education of Russian Federation within the program Priority-2030 (Strategic Technology Project of NUST MISIS "Quantum Internet") in part of numerical calculations.

\appendix

\section{Derivation of Eqs.~(\ref{eq:phi_m_hankel})-(\ref{eq:form_factor_slab}) in the main text}
\label{app:coulomb_closure}

Here we provide details of the derivation of Eqs.~(\ref{eq:phi_m_hankel})-(\ref{eq:form_factor_slab}) in the main text. The induced two-dimensional charge density is denoted by $\delta\rho(\mathbf r,t)$. The corresponding three-dimensional distribution $\delta \rho_{\rm 3D}$ entering the Poisson equation~(\ref{eq:poisson_3d}) is factorized as
\begin{equation}
    \delta\rho_{\rm 3D}(\mathbf r,z,t) = \delta\rho(\mathbf r,t)w(z) \ , \ \ \int dz w(z)=1 \ ,
\end{equation}
where $w(z)$ describes the transverse profile of the conducting layer. For the solution of Eq.~(\ref{eq:poisson_3d}) we use the angular expansions~(\ref{eq:angular_expansion_rho})-(\ref{eq:angular_expansion_Phi_3d}) and seek the solution for each angular harmonic in the Hankel representation
\begin{subequations}
\begin{align}
    \label{eq:app_rho_hankel}
    \delta\rho_m(r,t) = \int_0^\infty d\kappa \  \kappa \, \delta\rho_m(\kappa,t)J_m(\kappa r) \ ,\\
    \label{eq:app_phi_hankel}
    \Phi_{m}(r,t)  = \int_0^\infty d\kappa \ \kappa \, \Phi_{ m}(\kappa, t)J_m(\kappa r) \ ,\\
    \label{eq:app_phi_3d_hankel}
    \Phi_{{\rm 3D},m}(r,z,t)  = \int_0^\infty d\kappa \ \kappa \, \Phi_{{\rm 3D}, m}(\kappa,z,t)J_m(\kappa r) \ ,
\end{align}
\end{subequations}
The corresponding inverse transforms are
\begin{subequations}
\begin{align}
    \label{eq:app_rho_hankel_inverse}
    \delta\rho_m(\kappa,t) = \int_0^{R_s} dr \ r\, J_m(\kappa r)\delta\rho_m(r,t) \ ,\\
    \label{eq:app_phi_hankel_inverse}
    \Phi_m(\kappa,t) = \int_0^{R_s} dr \ r\, J_m(\kappa r)\Phi_m(r,t) \ ,\\
    \label{eq:app_phi_3d_hankel_inverse}
    \Phi_{{\rm 3D}, m}(\kappa, z, t) = \int_0^{R_s} dr \ r\, J_m(\kappa r)\Phi_{{\rm 3D}, m}(r,z, t) \ .
\end{align}
\end{subequations}
Substituting Eqs.~(\ref{eq:angular_expansion_rho}), (\ref{eq:angular_expansion_Phi_3d}) into Eq.~(\ref{eq:poisson_3d}) and using Eqs.~(\ref{eq:app_rho_hankel}) and (\ref{eq:app_phi_3d_hankel}), the Poisson equation is reduced to a one-dimensional equation in the transverse coordinate,
\begin{equation}
    \label{eq:app_z_equation}
    \left(\frac{\partial^2}{\partial z^2} - \kappa^2 \right) \Phi_{{\rm 3D}, m}(\kappa,z,t) = -4\pi \delta\rho_m(\kappa,t)w(z) \ ,
\end{equation}
and the solution decaying away from the 2D system has the form
\begin{equation}
    \label{eq:app_z_solution}
    \Phi_{{\rm 3D}, m}(\kappa,z,t) = \frac{2\pi}{\kappa} \delta\rho_m(\kappa,t) \int dz' e^{-\kappa|z-z'|} w(z') \ .
\end{equation}
Substituting this expression into Eq.~\eqref{eq:app_phi_3d_hankel} and then performing the transverse average by means of Eq.~(\ref{eq:phi_projected}), one gets
\begin{equation}
    \Phi_m(r,t) = 2\pi \int_0^\infty d\kappa \ F(\kappa) J_m(\kappa r) \delta\rho_m(\kappa,t) \ ,
    \label{eq:app_phi_kappa}
\end{equation}
where the finite-thickness form factor is defined by Eq.~(\ref{eq:form_factor_general}). Finally, using Eq.~\eqref{eq:app_rho_hankel_inverse}, the potential can be cast to the form presented by Eq.~\eqref{eq:phi_m_hankel} in the main text. For the numerical calculations we use a uniform-slab transverse profile,
\begin{equation}
   \label{eq:transverse_distribution_profile}
    w(z)=1/d \ , \ \ -d/2<z<d/2 \ .
\end{equation}
Substituting Eq.~(\ref{eq:transverse_distribution_profile}) into Eq.~(\ref{eq:form_factor_general}) and performing the integration, we obtain Eq.~(\ref{eq:form_factor_slab}) in the main text.

\section{Diffusive reduction of the Boltzmann equation}
\label{app:diffusive_reduction}

In this section we provide details of the derivation of basic equations in Sec.~\ref{sec:non_self_consistent} devoted to the discussion of the non-self-consistent approximation. We start from the linear response theory. Substituting Eq.~(\ref{eq:nsc_f1_expansion}) into Eq.~(\ref{eq:boltzmann}) with $\mathbf{E} = \mathbf{E}^{\rm ext}$ given by Eq.~(\ref{eq:E_real}), we get the linearized Boltzmann equation
\begin{equation}
    \left( -i\omega + \mathbf v\cdot\nabla_{\mathbf r}\right)f_\omega^{(1)}+e\mathbf E_\omega^{\rm ext}\cdot\nabla_{\mathbf p}f_0= -\frac{f_\omega^{(1)}-\langle f_\omega^{(1)}\rangle_{\mathbf p}}{\tau} \ .
    \label{eq:app_linear_boltzmann}
\end{equation}
In the diffusive limit the angular dependence of $f_\omega^{(1)}$ in the momentum space may be
truncated at the isotropic component and the first angular harmonic. Using the truncation~(\ref{eq:nsc_f1_truncation}) and projecting the resulting Eq.~(\ref{eq:app_linear_boltzmann}) onto harmonics gives
\begin{subequations}
\label{eq:app_nsc_first_order_harmonics_equations}
\begin{align}
     \label{eq:app_f0_linear_eq}
    -i\omega f_{\omega,0}^{(1)} + \frac{v}{2} \nabla_{\mathbf r}\cdot \mathbf f_{\omega,1}^{(1)} = 0 \ ,\\
    \mathbf{f}_{\omega,1}^{(1)} = -v\tau\nabla_{\mathbf{r}}f_{\omega,0}^{(1)} - v\tau e\mathbf{E}_{\omega}^{\rm ext}(\partial_\xi f_0) \ .
    \end{align}
\end{subequations}
To obtain the first equation in the above system one should average Eq.~(\ref{eq:app_linear_boltzmann}) over momentum directions, while the second equation can be obtained by multiplying  Eq.~(\ref{eq:app_linear_boltzmann}) by $(\hat{\mathbf{p}})_i$ ($i = x,y$) followed by momentum averaging. Finally, integrating Eqs.~(\ref{eq:app_nsc_first_order_harmonics_equations}) over the quasiparticle energy and using the expressions for the charge density modulation~(\ref{charge_density_definition}) and the current density~(\ref{current_density_definition}), one recovers Eqs.~(\ref{continuity_current_nsc}) in the main text.

We now turn to the second-order dc component. Substituting the second-order correction~(\ref{eq:nsc_f2_expansion}) into Eq.~(\ref{eq:boltzmann}) and collecting the time-independent terms in the Boltzmann equation, one gets
\begin{equation}
    \mathbf v\cdot\nabla_{\mathbf r} f_{\rm dc}^{(2)} + \left[ e\mathbf E_\omega^{\rm ext}\cdot\nabla_{\mathbf p} f_\omega^{(1)*} + {\rm c. c.} \right] = -\frac{f_{\rm dc}^{(2)}-\langle f_{\rm dc}^{(2)}\rangle_{\mathbf p}}{\tau}.
    \label{eq:app_second_order_dc_boltzmann}
\end{equation}
This equation is the starting point for the non-self-consistent second-order dc current. As in the linear problem, we keep only the isotropic and first angular harmonics [see Eq.~(\ref{eq:nsc_f2_truncation})].
The first angular harmonic is obtained by projecting
Eq.~\eqref{eq:app_second_order_dc_boltzmann} onto the anisotropic part, which yields
\begin{equation}
    \mathbf f_{{\rm dc},1}^{(2)} = -v\tau\nabla_{\mathbf r} f_{{\rm dc},0}^{(2)} - e\tau \left[ \mathbf E_\omega^{\rm ext}\frac{\partial f_{\omega,0}^{(1)*}}{\partial p}
    +{\rm c. c.} \right] \ .
    \label{eq:app_f2_dc_first_harmonic_vector}
\end{equation}
Substituting the second-order correction~(\ref{eq:nsc_f2_truncation}) with the above $\mathbf{f}_{{\rm dc},1}^{(2)}$ into the expression for the current density~(\ref{current_density_definition}) and performing the averaging over the momentum directions, we obtain
\begin{eqnarray}
\label{eq:nsc_jphi_preliminary}
\mathbf{j}_{{\rm dc}}^{(2)} = -e\tau \nu \int d\xi \, v^2 \nabla_{\mathbf{r}} f_{{\rm dc},0}^{(2)} \\
\nonumber
 - e^2\nu\tau \int d\xi \, v^2\left[\mathbf{E}_{\omega}^{\rm ext}\frac{\partial f_{\omega,0}^{(1)*}}{\partial \xi} + {\rm c. c.}\right] \ .
\end{eqnarray}
To derive Eq.~(\ref{eq:nsc_second_order_current_vector}) in the main text, we use the relation $v^2 = (2/m^*)(\xi + \mu)$ together with the expression for the first-order charge density modulation
\begin{equation}
\label{eq:charge_density_integral}
\delta \rho_{\omega}^{(1)}(\mathbf{r}) = 2e\nu \int d\xi \, f_{\omega,0}^{(1)}(\mathbf{r},\xi) \ .
\end{equation} 
In the first term of Eq.~\eqref{eq:nsc_jphi_preliminary}, $v^2$ is evaluated at the Fermi surface in the quasiclassical approximation. This gives the second-order diffusion current in Eq.~(\ref{eq:nsc_second_order_current_vector})
\begin{equation}
-e\nu\tau v_F^2\int d\xi\,\nabla_{\mathbf r}f_{\rm dc,0}^{(2)} = -D\nabla_{\mathbf r}\delta\rho_{\rm dc}^{(2)} .
\end{equation}
The second term should be treated before setting \(v=v_F\), because it
contains the derivative with respect to momentum. Performing the integration of the second term in Eq.~(\ref{eq:nsc_jphi_preliminary}) over $d\xi$ by parts 
\begin{equation}
\int d\xi\,v^2\frac{\partial f_{\omega,0}^{(1)}}{\partial \xi} = -\frac{2}{m^*}\int d\xi\,f_{\omega,0}^{(1)} \  ,
\end{equation}
and using Eq.~(\ref{eq:charge_density_integral}), one gets the bilinear source term in Eq.~(\ref{eq:nsc_second_order_current_vector}). Note that the absence of a boundary contribution in the above integral follows from the fact that the first-order solution $f_{\omega,0}^{(1)}\propto \partial f_0/\partial \xi$ is localized near the Fermi surface.

\section{Sampling of the Coulomb kernels}
\label{app:coulomb_sampling}

In this section we describe the numerical sampling of the Coulomb kernels entering the self-consistent first-order problem. For the $m=\pm 1$ angular sectors the induced field can be expressed in terms of two radial kernels. In dimensional variables we define
\begin{subequations}
\begin{align}
    \label{eq:app_K0_def}
    \mathcal{K}_0(r,r') = (q_{\rm TF}/2)v_F r' \int_0^\infty d\kappa \ \kappa F(\kappa)J_1(\kappa r')J_0(\kappa r) \ ,
    \\
    \label{eq:app_K2_def}
    \mathcal{K}_2(r,r') = (q_{\rm TF}/2)v_F r' \int_0^\infty d\kappa \kappa F(\kappa)J_1(\kappa r')J_2(\kappa r) \ .     
\end{align}
\end{subequations}
These kernels determine the two circular combinations of the induced electric field appearing in the kinetic equations. Direct numerical evaluation of Eqs.~\eqref{eq:app_K0_def} and \eqref{eq:app_K2_def} is inconvenient because the integrands are oscillatory and exhibit power-law decay. For this reason we substitute the explicit slab form factor~(\ref{eq:form_factor_slab}) into Eqs.~(\ref{eq:app_K0_def}), (\ref{eq:app_K2_def}) and separate the kernels into analytically known terms and exponentially convergent residual integrals.

Using the recurrence relations for Bessel functions~\cite{AbramowitzStegun1964} together with standard integrals involving Bessel functions~\cite{GradshteynRyzhik2007}, one obtains
\begin{subequations}
\begin{align}
	\label{eq:app_K0_sample}
    \mathcal{K}_0(r,r') = \frac{q_{\rm TF}v_F r'}{d^2}\bigg[ - I_0(r',r) +\frac{d}{r'}\Theta(r'-r) \ \ \ \ \ \  \\
    \nonumber
    +\frac{r'}{2} \int_0^\infty d\kappa \ e^{-\kappa d} \left[ J_0(\kappa r')+J_2(\kappa r')\right]J_0(\kappa r) \bigg] \ ,    
    \\
    \label{eq:app_K2_sample}
    \mathcal{K}_2(r,r') = \frac{q_{\rm TF}v_F r'}{d^2}\bigg[-I_2(r',r)+\frac{d r'}{r^2}\Theta(r-r') \ \ \ \ \\
    \nonumber
    +\frac{r'}{2}\int_0^\infty d\kappa \ e^{-\kappa d} \left[J_0(\kappa r')+J_2(\kappa r')\right]J_2(\kappa r)\bigg] \ .
\end{align}
\end{subequations}
Here we use the convention $\Theta(0)=1/2$. The remaining integrals in Eqs.~\eqref{eq:app_K0_sample} and \eqref{eq:app_K2_sample} are exponentially convergent and evaluated numerically. The non-exponential parts of the Bessel integrals  
\begin{subequations}
\label{eq:app_I0I2_def}
\begin{align}
    \mathcal{I}_0(r',r) = \int_0^\infty \frac{d\kappa}{\kappa} J_1(\kappa r')J_0(\kappa r) \ , \\
    \mathcal{I}_2(r',r) = \int_0^\infty \frac{d\kappa}{\kappa} J_1(\kappa r')J_2(\kappa r) \ ,
\end{align}
\end{subequations}
can be expressed through complete elliptic integrals~\cite{AbramowitzStegun1964}. We use the parameter convention for the complete elliptic integrals $K(m)$ and $E(m)$.

\noindent
For $r'<r$:
\begin{subequations}
\begin{align}
    \label{eq:app_I0_less}
    \mathcal{I}_0(r',r) = \frac{2r}{\pi r'} \\
    \nonumber
     \times \left[ E\left(\frac{r'^2}{r^2}\right)  - \left(1-\frac{r'^2}{r^2}\right)K\left(\frac{r'^2}{r^2}\right)\right] \ ,  \\
    \label{eq:app_I2_less}
    \mathcal{I}_2(r',r) = \frac{2}{3\pi r r'} \biggl[ (2r'^2-r^2) E\left(\frac{r'^2}{r^2}\right) \\
    \nonumber
    - (r'^2-r^2) K\left(\frac{r'^2}{r^2}\right) \biggl] \ .   
\end{align}
\end{subequations}
For $r'>r$:
\begin{subequations}
\begin{align}
    \label{eq:app_I0_greater}
    \mathcal{I}_0(r',r)= \frac{2}{\pi} E\left(\frac{r^2}{r'^2}\right) \ ,
    \\
    \label{eq:app_I2_greater}
    \mathcal{I}_2(r',r) = \frac{2}{3\pi r^2} \biggl[ (2r'^2-r^2) E\left(\frac{r^2}{r'^2}\right) \ \\
    \nonumber
    - 2(r'^2-r^2) K \left(\frac{r^2}{r'^2}\right) \biggl] \ .    
\end{align}
\end{subequations}
The limiting values at equal arguments are
\begin{equation}
    \mathcal{I}_0(r,r)=\frac{2}{\pi} \ , \ \ 
    \mathcal{I}_2(r,r)=\frac{2}{3\pi}.
    \label{eq:app_I_equal}
\end{equation}
These expressions are used to avoid loss of accuracy near $r=r'$, where the separate elliptic-integral terms contain nearly singular contributions but the full combinations remain continuous.

\section{Details of the numerical approach for the self-consistent linear problem}
\label{app:linear_solver}

In this section we describe the numerical approach used to solve the self-consistent linear problem. The first-order correction to the distribution function is expanded as
\begin{equation}
    f_{\omega}^{(1)} = -\frac{\partial f_0}{\partial \xi} \sum_{n=-n_{\rm max}}^{n_{\rm max}}\sum_{m = \pm 1} u_n^{(m)}(r)e^{im\varphi + in\theta} \ , 
    \label{eq:app_linear_expansion}
\end{equation}
where $n_{\rm max}$ is the maximum order of distribution function truncation in the momentum angular harmonics and  $\theta=\varphi_{\mathbf p}-\varphi$. In the expressions below we consider the case $n_{\rm max} = 2$.  Substituting the above expansion into the linearized Boltzmann equation gives Eq.~(\ref{eq:sc_linear_hierarchy}) in the main text. 
We first solve the linear problem for the $m=+1$ angular sector, driven by the amplitude $E_-$. In this
sector the external field is
\begin{equation}
    E_r^{{\rm ext}(+)}=E_- \ , \ \
    E_\varphi^{{\rm ext}(+)}=iE_- \  .
\end{equation}
The isotropic harmonic $u_0^{(+)}$ determines the induced charge density and,
therefore, the induced electrostatic field. Using the Coulomb kernels defined
in Appendix~\ref{app:coulomb_sampling}, the source terms can be written as
\begin{subequations}
\begin{align}
   \label{eq:app_S1_plus}
    S_1^{(+)}(r) = ev_F E_-  - (q_{\rm TF}/2)v_F\\
    \nonumber
    \times \int_0^\infty d\kappa \int_0^{R_s}dr'\,r'\kappa F(\kappa) J_1(\kappa r')J_0(\kappa r) u_0^{(+)}(r') \ ,
    \\
    \label{eq:app_Sm1_plus}
    S_{-1}^{(+)}(r) = (q_{\rm TF}/2)v_F\\
    \nonumber
    \times \int_0^\infty d\kappa \int_0^{R_s}dr'\,r'\kappa  F(\kappa) J_1(\kappa r')J_2(\kappa r) u_0^{(+)}(r') \ .
\end{align}
\end{subequations}
The first term in Eq.~\eqref{eq:app_S1_plus} is due to the external field,
whereas the remaining terms are generated by the self-consistent Coulomb
field.

For implementation of the collocation method~\cite{Trefethen2000_SpectralMethodsMATLAB} we introduce the dimensionless variables
\begin{equation}
    \rho=\frac{r}{R_s} \ , \ \ R=\frac{R_s}{\ell} \ , \ \ w=\omega\tau .
    \label{eq:app_dimensionless_vars}
\end{equation}
For the $|n| \leq 2$ truncation used in the numerical calculations, the regular
behavior at the origin is factored out as
\begin{subequations}
\label{eq:app_regular_variables}
\begin{align}
    u_2^{(+)}(r) = eE_-R_s\,\rho W_2(\rho) \ ,\\
    u_1^{(+)}(r) = eE_-R_s\,W_1(\rho) \ ,\\
    u_0^{(+)}(r) = eE_-R_s\,\rho W_0(\rho) \ ,\\
    u_{-1}^{(+)}(r) = eE_-R_s\,\rho^2 W_{-1}(\rho) \ ,\\
    u_{-2}^{(+)}(r) = eE_-R_s\,\rho^3 W_{-2}(\rho) \ ,
\end{align}
\end{subequations}
and the functions $W_n(\rho)$ are regular at $\rho=0$. With
\begin{equation}
    g=(1-iw)R,
    \qquad
    g_0=-iwR,
    \label{eq:app_g_defs}
\end{equation}
the dimensionless $|n|\leq2$ equations are
\begin{subequations}
\begin{align}
    \label{eq:app_eq_W0}
    g_0 \rho W_0 + \rho^2\frac{dW_{-1}}{dx} + \frac{dW_1}{dx} + 2\rho W_{-1} = 0 \ ,\\
    \label{eq:app_eq_W1}
    g W_1 + \rho\frac{dW_0}{dx} + \rho\frac{dW_2}{dx} + W_0+W_2 = 1 - \hat C_0 W_0 \ ,\\
    \label{eq:app_eq_Wm1}
    g\rho^2 W_{-1} + \rho^3\frac{dW_{-2}}{dx} + \rho\frac{dW_0}{dx} + 3\rho^2 W_{-2} = \hat C_2 W_0 \ ,\\
    \label{eq:app_eq_W2}
    g\rho W_2 + \frac{dW_1}{dx} = 0 \ ,\\
    \label{eq:app_eq_Wm2}
    g\rho^3 W_{-2} +\rho^2\frac{dW_{-1}}{dx} = 0 \ .
\end{align}
\end{subequations}
Here $x=2\rho-1$, and $\hat C_0,\hat C_2$ are the dimensionless Coulomb
operators. The terms $-\hat C_0W_0$ and $\hat C_2W_0$ are the two circular
components of the induced electric field.

The resulting Eqs.~(\ref{eq:app_eq_W0})-(\ref{eq:app_eq_Wm2}) are solved on the Chebyshev--Lobatto grid 
\begin{equation}
    x_j=\cos\frac{\pi j}{N_r} \ , \ \ \rho_j=\frac{1+x_j}{2} \ , \ \ j=0,\ldots,N_r \ .
    \label{eq:app_cheb_grid}
\end{equation}
Thus, $\rho_0=1$ corresponds to the disk edge and $\rho_{N_r}=0$ to the origin.
The differentiation matrix with respect to $x$ is denoted by $\hat D$. Its
entries are~\cite{Trefethen2000_SpectralMethodsMATLAB,WeidemanReddy2000_DifferentiationMatrixSuite}
\begin{equation}
D_{ij} =
\begin{cases}
\displaystyle
\frac{c_i}{c_j}
\frac{(-1)^{i+j}}{x_i-x_j} \ ,
& i\neq j \ ,
\\[1.0em]
\displaystyle
-\frac{x_i}{2(1-x_i^2)} \ ,
& i=j \ , \ \ 1\leq i\leq N_r-1 \ ,
\\[1.0em]
\displaystyle
\frac{2N_r^2+1}{6} \ ,
& i=j=0 \ ,
\\[1.0em]
\displaystyle
-\frac{2N_r^2+1}{6} \ ,
& i=j=N_r \ ,
\end{cases}
\label{eq:app_cheb_D}
\end{equation}
where
\begin{equation}
\label{cj_coefficients}
    c_j= \begin{cases} 2, & j=0,N_r \ ,\\ 1, & 1\leq j\leq N_r-1 \ . \end{cases}
\end{equation}
The radial integrals are evaluated using Clenshaw--Curtis quadrature on the
same grid. We write
\begin{equation}
    \int_0^1 d\rho\, f(\rho) \approx \sum_{j=0}^{N_r} q_j f(\rho_j) \ ,
    \label{eq:app_CC_rule}
\end{equation}
where the weights correspond to the interval $0\leq \rho\leq1$. They are
given by~\cite{ClenshawCurtis1960}	
\begin{equation}
    q_j = \frac{1}{N_r c_j} \sum_{k=0}^{\lfloor N_r/2\rfloor}\frac{2}{c_{2k}} \frac{\cos\left(2\pi j k/N_r\right)}{(1-4k^2)} \ .
    \label{eq:app_CC_weights}
\end{equation}
Here $\lfloor x \rfloor$ is the floor function, coefficients $c_j$ are defined by Eq.~(\ref{cj_coefficients}) and the same convention is used for $c_{2k}$. In particular, $c_{2k}=2$ when $2k=0$ or $2k=N_r$, and $c_{2k}=1$ otherwise. To write the system~(\ref{eq:app_eq_W0})-(\ref{eq:app_eq_Wm2}) in the matrix form, we introduce the solution vector 
\begin{equation}
    W_n=\left[ W_n(\rho_0), W_n(\rho_1), \ldots, W_n(\rho_{N_r}) \right]^T,
\end{equation}
and the following matrices
\begin{equation}
    \hat H=\mathrm{diag}(\rho_0,\rho_1,\ldots,\rho_{N_r}) \ , \ \ 
    \hat I=\mathrm{diag}(1,\ldots,1) \ .
\end{equation}
The Coulomb operators are represented by matrices $\hat C_0$ and $\hat C_2$. With the convention that the dimensionless kernels ${\mathcal K}_0$ and ${\cal K}_2$ include all radial-measure factors, their matrix elements are
\begin{equation}
    (\hat C_l)_{ij}= \frac{\lambda}{b^2} q_j\rho_j {\mathcal K}_l(\rho_i,\rho_j) \ , \ \   l =0,2 \ ,
    \label{eq:app_C_matrices}
\end{equation}
where
\begin{equation}
    \lambda=q_{\rm TF}R_s,
    \qquad
    b = d/R_s \ .
\end{equation}
The explicit construction of ${\mathcal K}_0$ and ${\mathcal K}_2$ is described in Appendix~\ref{app:coulomb_sampling}.

Before imposing the boundary and regularity conditions, the collocation system for the
$|n|\leq2$ truncation can be written as
\begin{widetext}
\begin{equation}
\begin{pmatrix}
g\hat H & \hat D & 0 & 0 & 0
\\
\hat H\hat D+\hat I & g\hat I
& \hat H\hat D+\hat I+\hat C_0 & 0 & 0
\\
0 & \hat D & g_0\hat H
& \hat H^2\hat D+2\hat H & 0
\\
0 & 0 & \hat H\hat D-\hat C_2
& g\hat H^2 & \hat H^3\hat D+3\hat H^2
\\
0 & 0 & 0 & \hat H^2\hat D & g\hat H^3
\end{pmatrix}
\begin{pmatrix}
W_2\\
W_1\\
W_0\\
W_{-1}\\
W_{-2}
\end{pmatrix}
=
\begin{pmatrix}
0\\
\mathbf 1\\
0\\
0\\
0
\end{pmatrix} \ .
\label{eq:app_linear_matrix_n2}
\end{equation}
\end{widetext}
The vector $\mathbf 1=(1,\ldots,1)^T$ represents the external $E_-$ drive.

The boundary conditions at the disk edge corresponding to specular reflection have the form
\begin{subequations}
\label{eq:app_edge_bc_W}
\begin{align}
    W_1(1)+W_{-1}(1)=0 \ ,\\
    W_2(1)-W_{-2}(1)=0 \ .
\end{align}
\end{subequations}
These conditions are imposed by replacing two rows of the matrix of the system~(\ref{eq:app_linear_matrix_n2}) at the point $j=0$. The origin is treated by using the regular variables in Eq.~\eqref{eq:app_regular_variables}. In practice, rows which become degenerate at $\rho=0$ are replaced by regularity conditions. A convenient choice is
\begin{subequations}
\label{eq:app_origin_bc}
\begin{align}
    \frac{dW_1}{dx}\biggl|_{\rho=0}=0 \ , \ \
    \frac{dW_0}{dx}\biggl|_{\rho=0}=0 \ ,\\
    \label{eq:app_origin_bc_Wm1Wm2}
    \frac{dW_{-1}}{dx}\biggl|_{\rho=0}=0 \ , \ \
    \frac{dW_{-2}}{dx}\biggl|_{\rho=0}=0 \ .
\end{align}
\end{subequations}
The derivatives in Eqs.~\eqref{eq:app_origin_bc} should be evaluated using the last row of the differentiation matrix, because $\rho=0$ corresponds to $j=N_r$.

For comparison we also use the lower $|n|\leq1$ truncation, obtained by setting $W_2=W_{-2}=0$ in Eq.~(\ref{eq:app_linear_matrix_n2}). The corresponding matrix problem is as follows:
\begin{equation}
\begin{pmatrix}
g\hat I & \hat H\hat D+\hat I+\hat C_0 & 0
\\
\hat D & g_0\hat H & \hat H^2\hat D+2\hat H
\\
0 & \hat H\hat D-\hat C_2 & g\hat H^2
\end{pmatrix}
\begin{pmatrix}
W_1\\
W_0\\
W_{-1}
\end{pmatrix}
=
\begin{pmatrix}
\mathbf 1\\
0\\
0
\end{pmatrix} \ .
\label{eq:app_linear_matrix_n1}
\end{equation}
The above system should be supplemented by the boundary condition at the disk edge
\begin{equation}
     \label{eq:app_n1_truncation_bc}
    W_1(1)+W_{-1}(1)=0 \ .
\end{equation}

After solving the $E_-$ sector, the harmonics $u_n^{(+)} = eE_-R_s\tilde{u}_n^{(+)}$ are then reconstructed in the following fashion:
\begin{subequations}
\label{eq:app_reconstruct_plus}
\begin{align}    
    \tilde u_2^{(+)} = \rho W_2 \ , \ \ \tilde u_1^{(+)} = W_1 \ , \ \
    \tilde u_0^{(+)} = \rho W_0 \ ,\\
    \tilde u_{-1}^{(+)} = \rho^2 W_{-1} \ , \ \
    \tilde u_{-2}^{(+)} = \rho^3 W_{-2} \ ,
\end{align}
\end{subequations}
and the total first-order electric field in this sector (normalized to $E_-$) is
\begin{subequations}
\label{eq:app_total_field_plus}
\begin{align}
    \tilde E_r^{(+)} = 1+(\hat C_2-\hat C_0)W_0 \ ,\\
    \tilde E_\varphi^{(+)} =  i-i(\hat C_2+\hat C_0)W_0 \ .
\end{align}
\end{subequations}
The remaining $m = -1$ sector (driven by $E_+$) does not need to be solved independently. It can be obtained from the $E_-$ solution by the following replacement of the angular harmonics:
\begin{equation}
    u_n^{(-)}(r;E_+) = u_{-n}^{(+)}(r;E_- \rightarrow E_+) \ .
    \label{eq:app_plus_minus_relation}
\end{equation}
Equivalently, if the regular amplitudes in the \(m=-1\) sector are defined by
\begin{subequations}
\begin{align}
u^{(-)}_2(r)=eE_+R_s\rho^3 W^{(-)}_2(\rho) \ ,\\
u^{(-)}_1(r)=eE_+R_s\rho^2 W^{(-)}_1(\rho) \ ,\\ 
u^{(-)}_0(r)=eE_+R_s\rho W^{(-)}_0(\rho) \ ,\\
u^{(-)}_{-1}(r)=eE_+R_s W^{(-)}_{-1}(\rho) \ ,\\
u^{(-)}_{-2}(r)=eE_+R_s\rho W^{(-)}_{-2}(\rho) \ ,
\end{align}
\end{subequations}
then Eq.~(\ref{eq:app_plus_minus_relation}) gives
\begin{subequations}
\begin{align}
W^{(-)}_2(\rho) = W^{(+)}_{-2}(\rho) \ ,\\
W^{(-)}_1(\rho) = W^{(+)}_{-1}(\rho) \ ,\\
W^{(-)}_0(\rho) = W^{(+)}_0(\rho) \ ,\\
W^{(-)}_{-1}(\rho) = W^{(+)}_1(\rho) \ ,\\
W^{(-)}_{-2}(\rho) = W^{(+)}_2(\rho) \ .
\end{align}
\end{subequations}
The corresponding projections of the total field in the $m = -1$ sector (normalized to $E_+$) are as follows:
\begin{subequations}
\label{eq:app_total_field_minus}
\begin{align}
    \tilde E_r^{(-)} = 1+(\hat C_2-\hat C_0)W_0^{(-)} \ ,\\
    \tilde E_\varphi^{(-)} = -i + i(\hat C_2+\hat C_0)W_0^{(-)} \ .    
\end{align}
\end{subequations}
Equations~\eqref{eq:app_linear_matrix_n2} and \eqref{eq:app_linear_matrix_n1}, supplemented by the boundary and regularity conditions~(\ref{eq:app_edge_bc_W}), (\ref{eq:app_origin_bc}), and (\ref{eq:app_n1_truncation_bc}) described above, define the self-consistent linear problem used to calculate the first-order observables and construct the second-order dc sources.

\begin{widetext}

\section{Details of the numerical approach for self-consistent calculations of the inverse Faraday effect}
\label{app:second_order_solver}

In this section we provide details of the numerical approach used for the self-consistent calculations of the inverse Faraday effect. Our starting point is the system~(\ref{second_order_dc_hierarchy}) in the main text. The source terms in these equations are given by 
\begin{subequations}
\label{eq:appE_momentum_sources}
\begin{align}
S_1 = -ie \,{\rm Im}\sum_{s = \pm 1}\left\{ - 2iE_{\varphi}^{(s)*}\partial_p F_0^{(s)} + V_s(r)\left(\partial_p + \frac{2}{p}\right)F_2^{(s)} - U_s(r)\left(\partial_p + \frac{2}{p}\right)F_{-2}^{(s)}\right\} \ ,\\
S_2 = -ie \, {\rm Im}\sum_{s = \pm 1}\left\{U_s\left(\partial_p - \frac{1}{p}\right)F_1^{(s)}- V_s(r)\left(\partial_p -\frac{1}{p}\right)F_{-1}^{(s)}\right\} \ ,\\
S_3 = -ie \, {\rm Im}\sum_{s = \pm 1}\left\{ U_s(r)\left(\partial_p - \frac{2}{p}\right)F_2^{(s)} - V_s(r)\left(\partial_p - \frac{2}{p}\right)F_{-2}^{(s)}\right\} \ .
\end{align}
\end{subequations}
Here $U_s(r) = E_r^{(s)*}-iE_{\varphi}^{(s)*}$ and $V_s(r) = E_r^{(s)*}+iE_{\varphi}^{(s)*}$. The crucial point in developing the self-consistent numerical approach for the inverse Faraday effect is that the momentum dependence in Eq.~\eqref{eq:appE_momentum_sources} has to be treated before performing the Fermi-surface approximation in the current. The same issue already appears in the non-self-consistent calculation [see, e.g., Appendix~\ref{app:diffusive_reduction} for details]. At zero temperature the momentum dependence of the distribution function harmonics is proportional to $\delta(\xi)$, while the source terms in Eqs.~(\ref{eq:appE_momentum_sources}) result in the appearance of contributions proportional to $\delta'(\xi)$. In particular, 
\begin{subequations}
\begin{align}
F_n^{(s)}(r,p)=-(\partial_\xi f_0)u_n^{(s)}(r)
=\delta(\xi)u_n^{(s)}(r) \ ,\\
\left( \partial_p+\frac{\alpha}{p} \right)F_n^{(s)} = v_Fu_n^{(s)}\delta'(\xi) + \frac{\alpha-1}{p_F}u_n^{(s)}\delta(\xi) \ .
\end{align}
\end{subequations}
Note that in order to obtain the second equation the relations $\partial_p=v\partial_\xi$, $v=v_F+\xi/p_F+\ldots$, and $\xi\delta'(\xi)=-\delta(\xi)$ were used. Thus, the numerical approach for the calculations of the inverse Faraday response requires the corresponding expansion of the solution and the source term
\begin{equation}
A_i = a_i\delta(\xi) + b_i\delta'(\xi) \ , \ \ 
S_i = s_i\delta(\xi) + t_i\delta'(\xi) \ , \ \ i = 1,2,3 \ .
\end{equation}
Here we drop the coordinate dependence of the amplitudes $a_i$, $b_i$, $s_i$, and $t_i$ for brevity. To construct the numerical scheme, we introduce the matrix differential operator
\begin{equation}
\hat{L}(r) = \frac{1}{2}\begin{bmatrix}0&(\partial_r + 2/r)&0\\ (\partial_r - 1/r)&0&(\partial_r + 3/r)\\0&(\partial_r - 2/r)&0\end{bmatrix} \ ,
\end{equation}
and the column vectors $\mathbf{a} = [a_1,a_2,a_3]^{\rm T}$, $\mathbf{b} = [b_1,b_2,b_3]^{\rm T}$, $\mathbf{s} = [s_1,s_2,s_3]^{\rm T}$, and $\mathbf{t} = [t_1,t_2,t_3]^{\rm T}$ containing all the solution and the source components. As a result, one obtains the two linear systems
\begin{subequations}
\begin{align}
\tau^{-1}\mathbf{a} + v_F\hat{L}\mathbf{a}  = \mathbf{s} + \frac{1}{p_F}\hat{L}\mathbf{b} \ ,\\
\tau^{-1}\mathbf{b} + v_F\hat{L}\mathbf{b} = \mathbf{t} \ . 
\end{align}
\end{subequations}
Using Eq.~(\ref{eq:app_regular_variables}) and Eqs.~(\ref{eq:app_reconstruct_plus})-(\ref{eq:app_total_field_minus}), we can rewrite the source terms using dimensionless electric field projections and distribution function harmonics
\begin{subequations}
\begin{align}
s_1 = \frac{-ie^2R_s E_0^2}{p_F} {\rm Im}\biggl\{\frac{|E_-|^2}{E_0^2}\left[2i\tilde{E}_{\varphi}^{(+)*}\tilde{u}_0^{(+)}+\tilde{V}_+\tilde{u}_2^{(+)}-\tilde{U}_+\tilde{u}_{-2}^{(+)}\right] \\
\nonumber
 + \frac{|E_+|^2}{E_0^2}\left[2i\tilde{E}_{\varphi}^{(-)*}\tilde{u}_0^{(-)}+\tilde{V}_-\tilde{u}_2^{(-)}-\tilde{U}_-\tilde{u}_{-2}^{(-)}\right] \biggr\} \ ,\\
s_2 = \frac{-2ie^2R_sE_0^2}{p_F}{\rm Im}\left\{\frac{|E_-|^2}{E_0^2}\left[-\tilde{U}_+\tilde{u}_1^{(+)} + \tilde{V}_+\tilde{u}_{-1}^{(+)}\right] + \frac{|E_+|^2}{E_0^2}\left[-\tilde{U}_-\tilde{u}_1^{(-)} + \tilde{V}_-\tilde{u}_{-1}^{(-)}\right] \right\} \ ,\\
s_3 = \frac{-3ie^2R_sE_0^2}{p_F}{\rm Im}\left\{\frac{|E_-|^2}{E_0^2}\left[-\tilde{U}_+\tilde{u}_2^{(+)} + \tilde{V}_+\tilde{u}_{-2}^{(+)}\right] + \frac{|E_+|^2}{E_0^2}\left[-\tilde{U}_-\tilde{u}_2^{(-)} + \tilde{V}_-\tilde{u}_{-2}^{(-)}\right] \right\} \ ,\\
t_1 = -ie^2v_FR_sE_0^2{\rm Im}\biggl\{\frac{|E_-|^2}{E_0^2}\left[-2i\tilde{E}_{\varphi}^{(+)*}\tilde{u}_0^{(+)} + \tilde{V}_+\tilde{u}_2^{(+)}-\tilde{U}_+\tilde{u}_{-2}^{(+)}\right] \\
\nonumber
 + \frac{|E_+|^2}{E_0^2}\left[-2i\tilde{E}_{\varphi}^{(-)*}\tilde{u}_0^{(-)} + \tilde{V}_-\tilde{u}_2^{(-)}-\tilde{U}_-\tilde{u}_{-2}^{(-)}\right]\biggr\} \ ,\\
t_2 = -ie^2v_FR_sE_0^2{\rm Im}\left\{\frac{|E_-|^2}{E_0^2}\left[\tilde{U}_+\tilde{u}_1^{(+)}-\tilde{V}_+\tilde{u}_{-1}^{(+)}\right] + \frac{|E_+|^2}{E_0^2}\left[\tilde{U}_-\tilde{u}_1^{(-)}-\tilde{V}_-\tilde{u}_{-1}^{(-)}\right]\right\} \ ,\\
t_3 = -ie^2v_FR_sE_0^2{\rm Im}\left\{\frac{|E_-|^2}{E_0^2}\left[\tilde{U}_+\tilde{u}_2^{(+)}-\tilde{V}_+\tilde{u}_{-2}^{(+)}\right] + \frac{|E_+|^2}{E_0^2}\left[\tilde{U}_-\tilde{u}_2^{(-)}-\tilde{V}_-\tilde{u}_{-2}^{(-)}\right]\right\} \ .
\end{align}
\end{subequations}
Here $\tilde U_s=\tilde E_r^{(s)*}-i\tilde E_\varphi^{(s)*}$ and $\tilde V_s=\tilde E_r^{(s)*}+i\tilde E_\varphi^{(s)*}$. The total first-order field in the $s=+1$ sector is normalized to $E_-$, whereas the field in the $s=-1$ sector is normalized to $E_+$. As a next step, we put $r = R_s\rho$, introduce the dimensionless solution and source components: $\tilde{\mathbf{a}} = \mathbf{a}/(e^2R_s\tau E_0^2/p_F)$, $\tilde{\mathbf{b}} = \mathbf{b}/(\tau e^2 v_FR_s E_0^2)$, $\tilde{\mathbf{s}} = \mathbf{s} p_F/(e^2R_sE_0^2)$, and $\tilde{\mathbf{t}} = \mathbf{t}/(e^2v_FR_sE_0^2)$. The dimensionless differential operator is defined as $\hat{\tilde{L}} = \hat{L}R_s$. For this choice we obtain the following system of equations:
\begin{subequations}
\begin{align}
\tilde{\mathbf{b}} + \frac{1}{R}\hat{\tilde{L}}\tilde{\mathbf{b}} = \tilde{\mathbf{t}} \ ,\\
\tilde{\mathbf{a}} + \frac{1}{R}\hat{\tilde{L}}\tilde{\mathbf{a}} = \tilde{\mathbf{s}} + \frac{1}{R}\hat{\tilde{L}}\tilde{\mathbf{b}} \ .
\end{align}
\end{subequations}
Here $R = R_s/\ell$. For implementation of the collocation method we factorize the regular behavior of the solution components at the origin 
\begin{equation}
\tilde{a}_i = \rho^i \alpha_i \ , \ \ \tilde{b}_i = \rho^i \beta_i \ , 
\end{equation}
and obtain the following linear systems:
\begin{equation}\label{second_order_numerical_system}
\begin{bmatrix}\rho&0&0\\ 0&\rho^2&0\\ 0&0&\rho^3\end{bmatrix}\begin{bmatrix}\beta_1\\ \beta_2\\ \beta_3\end{bmatrix} + \frac{1}{2R}\hat{L}_{\rho}\boldsymbol{\beta} = \tilde{\mathbf{t}} \ , \ \ \ 
\begin{bmatrix}\rho&0&0\\0&\rho^2&0\\0&0&\rho^3\end{bmatrix}\begin{bmatrix}\alpha_1\\ \alpha_2\\ \alpha_3\end{bmatrix} + \frac{1}{2R}\hat{L}_{\rho}\boldsymbol{\alpha} = \tilde{\mathbf{s}} + \frac{1}{2R}\hat{L}_{\rho}\boldsymbol{\beta} \ ,
\end{equation}
where $\boldsymbol{\alpha} = [\alpha_1,\alpha_2,\alpha_3]^{\rm T}$, $\boldsymbol{\beta} = [\beta_1,\beta_2,\beta_3]^{\rm T}$, and
\begin{equation}
\hat{L}_{\rho} = \begin{bmatrix}0&\rho^2\partial_{\rho} + 4\rho&0\\ \rho\partial_{\rho}&0&\rho^3\partial_{\rho} + 6\rho^2\\ 0&\rho^2\partial_{\rho}&0\end{bmatrix} \ .
\end{equation}
Both systems in Eqs.~(\ref{second_order_numerical_system}) are solved on the Chebyshev--Lobatto grid introduced in Appendix~\ref{app:linear_solver}. The boundary condition at the disk edge follows from specular
reflection and gives
\begin{equation}
A_2(R_s,p)=0 \ .
\end{equation}
Since $A_2=a_2\delta+b_2\delta'$, this condition is imposed as
\begin{equation}
\alpha_2(1)=0 \ , \ \  \beta_2(1)=0 \ .
\end{equation}
At the origin the functions $\alpha_i$ and $\beta_i$ are regular. In the collocation implementation the rows that become degenerate at \(\rho=0\) are replaced by regularity conditions for the corresponding regular amplitudes $\partial_\rho \alpha_i|_{\rho = 0} = 0$ and $\partial_{\rho}\beta_i|_{\rho = 0} = 0$. Finally, the circulating current is reconstructed from its original velocity-weighted definition,
\begin{equation}
j_\varphi^{(2)}(r) = ie \int\frac{pdp}{2\pi} v(p)A_1(r,p) \ .
\end{equation}
Using $A_1=a_1\delta+b_1\delta'$, one obtains
\begin{equation}
j_\varphi^{(2)}(r) =
ie\nu \left[ v_Fa_1(r)-\frac{1}{p_F}b_1(r) \right] \ .
\end{equation}
In dimensionless form this gives
\begin{equation}
j_\varphi^{(2)}(\rho) = \frac{ie^3\nu\tau R_sE_0^2}{m^*}\, \rho\left[ \alpha_1(\rho)-\beta_1(\rho) \right] \ .
\label{eq:appE_current_final}
\end{equation}
This expression is used to compute the area-integrated current and the magnetic moment from Eqs.~(\ref{eq:Mz_def}) and~(\ref{eq:Iarea_def}) in the main text.

\end{widetext}

\bibliography{bibliography}

\end{document}